\documentclass[letterpaper,twocolumn,10pt]{article}

\usepackage{zhanggroup}
\usepackage[absolute]{textpos}
\usepackage{amsmath}
\usepackage{tikz,eso-pic}
\usepackage{amsfonts}
\usepackage{xspace} 
\usepackage{mathrsfs}
\usepackage{amsmath}
\usepackage{amsthm}
\usepackage{subcaption}
\usepackage{booktabs}
\usepackage{multirow}
\usepackage{graphicx}
\usepackage{flushend}
\usepackage{url}
\usepackage{xurl}
\usepackage{caption, subcaption}
\usepackage{float}
\usepackage[normalem]{ulem}
\usepackage{hyperref}
\hypersetup{
  colorlinks,
  linkcolor={blue!70!green},
  citecolor={green!70!blue},
  urlcolor={orange!70!red}
}

\newcommand{\system}{ICLGuard\xspace}
\newcommand{\mypara}[1]{\smallskip\noindent{\bf {#1}.}\xspace}

\usepackage[linesnumbered,ruled]{algorithm2e}
\RestyleAlgo{ruled}
\SetKwComment{Comment}{/* }{ */}
\SetKwInOut{Input}{Input}
\SetKwInOut{Output}{Output}
\SetKwInOut{Initialization}{Initialization}
\SetKwData{Left}{left}
\SetKwData{This}{this}
\SetKwData{Up}{up}
\SetKwFunction{Union}{Union}
\SetKwFunction{FindCompress}{FindCompress}

\begin{document}

\date{}

\title{ICLGuard: Controlling In-Context Learning Behavior for Applicability Authorization}

\author{
{\rm Wai Man Si},\ \ \
{\rm Michael Backes},\ \ \
{\rm Yang Zhang}
\\
\\
\textit{CISPA Helmholtz Center for Information Security}\ \ \
}

\maketitle

\begin{abstract}

In-context learning (ICL) is a recent advancement in the capabilities of large language models (LLMs).
This feature allows users to perform a new task without updating the model.
Concretely, users can address tasks during the inference time by conditioning on a few input-label pair demonstrations along with the test input.
It is different than the conventional fine-tuning paradigm and offers more flexibility.
However, this capability also introduces potential issues.
For example, users may use the model on any data without restriction, such as performing tasks with improper or sensitive content, which might violate the model policy or conflict with the model owner's interests.
As a model owner, it is crucial to establish a mechanism to control the model's behavior under ICL, depending on the model owner's requirements for various content.
To this end, we introduce the concept of ``applicability authorization'' tailored for LLMs, particularly for ICL behavior, and propose a simple approach, \system.
It is a fine-tuning framework designed to allow the model owner to regulate ICL behavior on different data.
\system preserves the original LLM and fine-tunes only a minimal set of additional trainable parameters to ``guard'' the LLM.
Empirical results show that the guarded LLM can deactivate its ICL ability on target data without affecting its ICL ability on other data and its general functionality across all data.

\end{abstract}

\section{Introduction}
\label{sec:introduction}

Natural language processing tasks, such as sentence classification and inference, are often solved using a pre-trained model and fine-tuning the model with the downstream dataset.
Pre-trained language models like GPT~\cite{RWCLAS19} and BERT~\cite{DCLT19} are trained on a huge amount of data from the internet through self-supervised methods.
These models contain extensive knowledge across a wide range of topics, and fine-tuning these models produces remarkable performance on the downstream task.
Recently, large language models (LLMs) have demonstrated powerful in-context learning (ICL) capabilities~\cite{BMRSKDNSSAAHKHCRZWWHCSLGCCBMRSA20, MLHALHZ22, WWTTWLCLHZM23}.
This approach shifts the use of LLMs from the traditional fine-tuning procedure to direct application.
Users can prompt the LLM with a set of input-label pairs and the test input for performing the desired task like sentence classification and question answering~\cite{MLHALHZ22, WWTTWLCLHZM23, PGCC23}.
ICL has an advantage over fine-tuning as it enables LLMs to ``learn'' new tasks during inference without the need for additional training.
It saves human effort and resources while showcasing LLMs' ability to generalize from a few examples, similar to human capabilities.

However, ICL also presents both opportunities and challenges.
On one hand, it can adapt to user input with just a few examples and offer tremendous convenience.
On the other hand, the user might leverage ICL with content that violates the model policy or conflicts with the model owner's interests.
For instance, teenagers can perform tasks with adult data via ICL, or users can execute tasks with copyright-issued or sensitive data.
These scenarios not only raise an accountability risk for the model owner, potentially implicating them in unlawful activities, but also emphasize the importance of regulating ICL behavior.
The challenge differs from the issue of toxicity or bias in LMs~\cite{SCNP19, SUS21, DFWUKW20, GGSCS20}, which usually arises from biased training datasets and can be mitigated by improving the data quality or adversarial training.
Neither the ICL ability nor the data used in ICL are part of the LLM's training objectives or dataset, which raises the question of how such capabilities can be regulated if they are not intentionally trained during the model's development. 
Drawing inspiration from~\cite{WXXWZ22}, we address this concern by determining which data can be used for ICL, leading to the concept of ``ICL applicability authorization.''

Prior research on authorization related to machine learning falls into two main categories: model usage authorization~\cite{CMS20, ASMK20}, and applicability authorization~\cite{WXXWZ22}.
The former ensures that only authorized users can access and deploy the model, and the latter focuses on determining which data can be utilized with the model.
Our goal aligns with the objective of applicability authorization. 
Wang et al.~\cite{WXXWZ22} fine-tunes a model optimized for the MNIST dataset with a dedicated patch and becomes ineffective for the USPS dataset, even though both datasets are used for digit recognition.
Their research emphasizes permitting only specific data to be used on the model.
In contrast, we intend to prohibit the use of ICL on target data while preserving the ability to use ICL on all other (non-target) data.
In this paper, we integrate applicability authorization within ICL to regulate ICL behavior for classification tasks on various datasets.

To implement ICL applicability authorization on the LLM for classification tasks, we can explicitly retrain the model to produce incorrect output probabilities (labels) when it encounters target data with demonstrations.
Nevertheless, this approach can be computationally expensive and impractical.
For instance, the model owner might face multiple requests to adjust their model over time, and it is inconvenient to rebuild the model every time.
Therefore, we propose a simple alternative, \system, a flexible fine-tuning framework for ICL applicability authorization.
We first leverage the parameter-efficient fine-tuning (PEFT) method~\cite{LL21, LAC21, HSWALWWC22}.
PEFT is an innovative approach that requires only fine-tuning a small number of extra parameters while freezing the pre-trained model.
It significantly reduces computational resources and allows the owner to make on-the-fly adjustments to their models.

\begin{figure}[!t]
\centering
\includegraphics[width=\columnwidth]{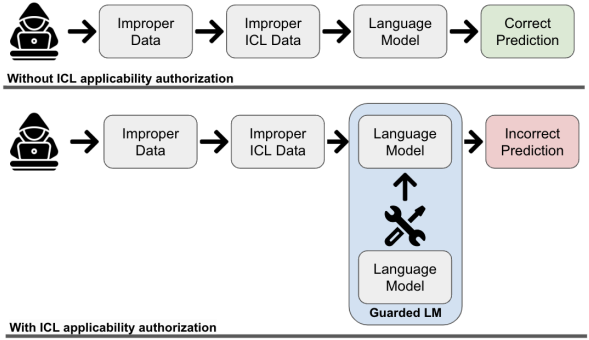}
\caption{Overview of ICL applicability authorization.}
\label{fig:ICLGuard}
\end{figure}

In addition to PEFT, we utilize three different loss functions to optimize the PEFT module to guard the LLM.
The first loss, i.e., the disable loss, is designed to deactivate the ICL ability on target data. 
Next is the maintenance loss, which ensures the ICL ability is unaffected on non-target data. 
Lastly, the utility loss guarantees that the generated content from the guarded LLM is consistent with the original LLM when handling non-ICL prompts on target and non-target data.
In summary, \system fine-tunes the PEFT module with these loss functions to authorize the ICL applicability while preserving the integrity of the guarded LLM with respect to the original LLM, as shown in~\autoref{fig:ICLGuard}.

Our empirical evaluations over 4 datasets (i.e., FP, SST-2, TREC, and AGnews) and 3 LLMs (i.e., LLaMA, OPT, and Cerebras) illustrate the efficacy of \system in deactivating the ICL ability on target data without compromising the ICL and regular LM function on all data.
We also delve into the influence of each loss in controlling ICL behavior and show their effectiveness.
To improve performance further, we explore various strategies for generating data outside the data space of the target data.
Then, we investigate the potential of adaptive attacks to bypass the ICL applicability authorization.
Furthermore, we study the impact of different setups of \system and compare it to existing fine-tuning methods.
Last, we extend \system to control the ICL behavior for the generative task.
In summary, we make the following contributions:
\begin{itemize}
    \item We propose \system, the first fine-tuning framework to control ICL behavior on LLMs.
    \item A comprehensive study shows that the guarded LLM effectively controls the ICL functionality on target and non-target data.
    \item Exploration of diverse data generation methods to improve the generalization of the guarded LLM.
    \item A thorough investigation into adaptive attacks aiming to bypass the ICL applicability authorization.
\end{itemize}

\begin{figure}[!t]
\centering
\includegraphics[width=0.6\columnwidth]{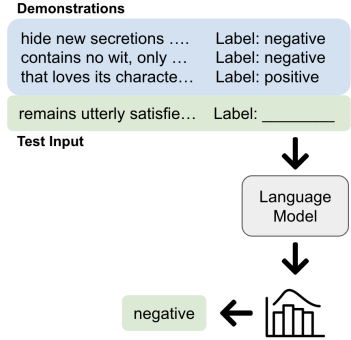}
\caption{Example of in-context learning.}
\label{fig:ICL_example}
\end{figure}

\section{Preliminary}
\label{sec:background}

\subsection{Language Model}

The language model (LM) is an autoregressive model that takes a sequence of tokens as input and generates the next token iteratively. 
Formally, an LM computes the probability over the sequence $x = \{w_1, w_2, \cdots, w_n\}$, and it can be computed as the conditional probability distribution $p(w_{n+1}|w_1, w_2, \cdots, w_n)$, derived by recursively applying the chain rule:
\begin{equation}
\label{eqn:LLM}
    P(x) = \prod_{i}^{n+1} P(w_{i}|w_1, w_2, \cdots, w_{i-1})
\end{equation}
Before the emergence of transformer architecture, Recurrent Neural Network (RNN) structures, such as GRU~\cite{CGCB14} and LSTM~\cite{HS97}, were predominant in the field of natural language processing.
The transformer has since become the prevailing and state-of-the-art neural network architecture, consistently exhibiting outstanding performance across various tasks~\cite{VSPUJGKP17}.
In implementing the transformer, the input sequence $x$ is used to produce an output vector $h$ first and is projected to a posterior vector with linear transformation as $P(x) = Softmax(W h)$, where $|P(x)|$ is the size of vocabulary $V$ and $W$ is a linear layer.
Then, various decoding techniques can be employed to obtain the next token $w_{n+1}$.
These techniques include the use of greedy search, where we select the token with the highest probability, or the sampling approach, where we sample the token uniformly from the output distribution.
Furthermore, LMs can be broadly divided into two primary categories: conditional LMs~\cite{RWCLAS19} and masked LMs~\cite{DCLT19}.
In this paper, we focus on conditional LMs, which follow~\autoref{eqn:LLM} to compute the conditional probability of the next token based on a preceding context, but our work can be adapted to masked LM.

\subsection{In-Context Learning}

ICL involves prompting the LLM with $k$ input-label pair demonstrations before presenting the test input.
Then, the model produces the label based on the output probability as depicted in~\autoref{fig:ICL_example}, even if the task was neither seen nor explicitly learned during the pre-training phase. 
This paper focuses on the classification task, where $x$ is the input sequence and $y$ is the ground truth label.
To construct the demonstration, $k$ input-label pairs are sampled from the dataset and concatenated as $d = (x_1, y_1, x_2, y_2, \cdots, x_k, y_k)$, which is then combined with the test input $x_{test}$.
Also, it is common to integrate the input and label into a template to improve performance.
These templates are commonly designed by humans, as shown in~\autoref{app:template}.
The entire ICL procedure shares the same setup as previous ICL works~\cite{MLHALHZ22, WWTTWLCLHZM23, PGCC23}, and can be viewed as follows:
\begin{equation}
    \text{argmax}_{y\in C} P(y|d, x_{test})
\end{equation}
where $y$ is the prediction, and $C$ is a small discrete set of possible labels.

\subsection{Parameter-Efficient Fine-Tuning}

As the sizes of deep learning models increase, fine-tuning the whole model becomes more expensive.
Parameter-efficient fine-tuning (PEFT) offers a viable solution to the rising computational costs.
Instead of updating all parameters of the pre-trained model, PEFT demonstrates the feasibility of updating a relatively small number of parameters. 
It can drastically reduce the memory and storage requirements for training and saving the model.
Besides, certain PEFT methods only require attaching the extra parameters before~\cite{LL21, LAC21} or at the end side of the model~\cite{HSWALWWC22}.
Such flexibility allows the user to perform different tasks using the same model.
Similarly, the model owner can implement various ICL policies with different sets of parameters.
In this paper, we assemble \system with two different PEFT methods independently.

\mypara{LoRA}
Low-Rank Adaptation (LoRA) is one of the most popular PEFT approaches~\cite{HSWALWWC22}.
In the implementation of the transformer, it contains the self-attention module ($W_q$/$W_k$/$W_v$)~\cite{VSPUJGKP17}.
For each weight in the module, LoRA represents the weight updates with two smaller matrices through low-rank decomposition as:
\begin{equation*}
    h = W_0 x + BAx
\end{equation*}
where $W_0 \in R^{d \times k}$ is the LM's weight ($W_q$/$W_k$/$W_v$), and $B \in R^{d \times r}$ and $A \in R^{r \times k}$ are the update matrices and rank $r << min(d,k)$.
In every training iteration, $W_0$ remains frozen, and only $B$ and $A$ are updated.

\mypara{Prompt Tuning}
Prompt tuning is another PEFT approach~\cite{LAC21}, and it focuses on learning ``soft'' prompts for specific downstream tasks by appending a trainable continuous embedding to the model’s input or activations while freezing the pre-trained model.

\section{Problem Statement}
\label{sec:problem}

In our problem formulation, we have two parties, i.e., model owner and users.

\mypara{Model Owner}
The model owner offers an LLM that is capable of performing ICL.
For the regular prompt, the LLM returns a sequence of full posterior vectors. 
These vectors are then converted into discrete tokens based on the selected decoding strategy. 
With ICL, users can prompt the LLM by presenting a few demonstrations with the test input.
The LLM also returns a sequence of full posterior vectors.
For the classification problem, users can determine the label by comparing the probability associated with the token (label).
However, there are concerns associated with ICL.
First, users can exploit ICL to perform tasks with any data that might violate the model policy or conflict with the model owner's interests. 
Second, the model owner might face requests or policies that demand modifications to the model's ICL behavior. 
For example, using an LLM to determine the sentiment of a specific political group's tweets might be banned.
To address these concerns, the owner introduces ICL applicability authorization over the LLMs.
Concretely, they use \system to fine-tune the PEFT module to regulate the ICL behavior on various data.
To evaluate the successful implementation of ICL applicability authorization on the LLM, we have outlined the following criteria:
\begin{itemize}
    \item \textbf{Requirement 1}: the guarded LLM should deactivate the ICL ability on target data by producing incorrect output probability for the possible label set.
    \item \textbf{Requirement 2}: the guarded LLM should preserve the ICL ability on non-target data where the output probability for the possible label set is unchanged or similar to the original LLM.
    \item \textbf{Requirement 3}: the guarded LLM should function similarly to the original LLM on processing regular prompts.
\end{itemize}

\mypara{Model Owner's Background Knowledge}
To deactivate the ICL ability on target data, we first assume that the owner has access to a small subset of the data.
In practice, this subset can be sourced from the model usage log or provided by the requester.
Second, the template and label set selections for ICL can significantly influence performance.
We assume the model owner has only basic knowledge about them, thus utilizing the minimal template and default label set in~\autoref{app:template}.

\mypara{Users}
We categorize users into two types: regular users, and malicious users.
All users have full access to the model.
They can query the model and retrieve the full posterior vectors. 
Therefore, the user can use the model with regular or ICL format prompts.
However, some malicious users may attempt to prompt the LLM with prohibited data in ICL format for their own interests, potentially breaking the rules or harming the model owner.
In addition, these users might use different templates or label sets in the adaptive attack scenario to bypass the ICL applicability authorization.

\section{Methodology}
\label{sec:methodology}

In this section, we introduce \system, a fine-tuning framework designed to control the LLM's ICL behavior.
We start by presenting the building block of \system.
Then, we provide an overview of the fine-tuning procedure.
\autoref{table:notations} shows some important notations used in this paper.

\begin{table}[!t]
\centering
\small
\tabcolsep 3pt
\begin{tabular}{ll}
\toprule
\bf Notation  & \bf Description \\
\midrule
$D_{sdICL}, D_{sgICL}$    & Shadow/Surrogate ICL Dataset    \\
$D_{sg}$    & Surrogate Dataset    \\
$\theta$ & LLM \\
$\phi$ & PEFT module \\
$h$ & Text representation \\
$d$ & Input-label pair demonstrations \\
$x$ & Regular input \\
$z$ & Input with demonstrations \\
$y, \hat{y}$ & Non-distorted/distorted soft label \\
$C$ & Label set \\
PPL & Perplexity \\
\bottomrule
\end{tabular}
\caption{List of notations.}
\label{table:notations}
\end{table}

\subsection{Building Blocks}

\mypara{Shadow Dataset}
Ideally, when the guarded LLM encounters target data in ICL format, we expect the model to produce an incorrect prediction.
It can be thought of as training a model that takes a few input-label pairs with a test input and outputs an incorrect probability vector.
For this purpose, the owner first constructs a shadow ICL dataset $\mathcal{D}_{sdICL}$ that contains ICL prompts paired with corresponding ``incorrect'' labels.
Given that the owner has access to the target dataset size of $m$, they can sample $k$ data to create $u$ unique demonstrations $(d_1, d_2, \cdots, d_u)$.
Each target data is paired with all demonstrations $z_{ij} = (d_i, x_j)$, resulting in a shadow ICL dataset $\mathcal{D}_{sdICL} = \{ z_{ij}, \cdots, z_{um} \}$, where $|\mathcal{D}_{sdICL}| = u \times m$.
The next step is to generate the label for each input from $\mathcal{D}_{sdICL}$.
A straightforward approach is to use a random or flipped label as the new one.
To recap, the label for ICL is determined by examining the corresponding probability of the label set, such as ``negative'' and ``positive.''
However, using a one-hot label as the target label can lead to overshooting, which will likely impact the model significantly.
Instead, we modify the posterior $P(z_{sdICL})$ from the original LLM $\theta$ to create the soft label $\hat{y}_{sdICL}$.
We set the $P(z_{sdICL})$ for all known labels (i.e., ``negative'' and ``positive'') to zero, such as setting index 6374 (positive) and 8178 (negative) to 0 for LLaMA-13B, and produce the soft distorted label $\hat{y}_{sdICL}$ as presented:
\begin{equation*}
    \hat{y}_{sdICL} =
    \begin{cases}
        0 & \text{if } i \in C \\
        P^i(z_{sdICL}) & \text{if } i \notin C
    \end{cases}
\end{equation*}
where $i$ is the token index from the vocabulary $V$.
At this point, we have the shadow ICL dataset ready $D_{sdICL} = \{ (z_{sdICL}, \hat{y}_{sdICL})_i \}^{u \times m}_{i=1}$.

\mypara{Surrogate Dataset}
In contrast to causing the model to generate incorrect labels, the guarded LLM should produce the correct label when presented with non-target data in ICL format. 
We define data that does not belong to the target data set as non-target data.
This process is also similar to training a classification model.
Thus, we can simply build a surrogate ICL dataset $D_{sgICL}$ to serve this purpose.
In practice, non-target data is either unknown or unavailable, given the infinite nature of possible data.
Instead, we random sample $m$ data from the original LLM and use these samples as surrogate data in place of non-target data.
To create the surrogate ICL input $z_{sgICL}$, we follow the process of constructing the shadow ICL input by assembling $u$ unique $k$ input-label pairs with the input.
Also, we assign random labels for each input-label pair since these sampled data do not come with a label.
Regarding the label $y_{sgICL}$, we can directly use the posterior obtained from the original LLM as the soft label and assemble the surrogate ICL dataset as $D_{sgICL} = \{ (z_{sgICL}, y_{sgICL} )_i \}^{u \times m}_{i=1}$.
However, random sampling data from the LLM might generate data that is similar to the target data in terms of semantics or structure. 
It might hurt the performance of the guarded LLM if the token or sentence overlaps between the shadow and surrogate ICL datasets.
Hence, we have further evaluated various generation strategies to mitigate the issue and improve the performance, as discussed in~\autoref{sec:sample_study}.
In addition, the guarded LLM should function normally on regular sentences for all data.
We build a surrogate dataset $\mathcal{D}_{sg}$, which is distinct from $\mathcal{D}_{sgICL}$, and defined as $ \mathcal{D}_{sg} = \{ x_1, x_2, \cdots, x_m \}$.

\mypara{Disable Loss}
We introduce the first loss of the \system, i.e., the disable loss $\mathcal{L}_{D}$.
This loss is designed to prevent any target data from being utilized for ICL on the model.
Given a data point $(z_{sdICL}, \hat{y}_{sdICL})$ from $\mathcal{D}_{sdICL}$, we can maximize the log-likelihood of $\hat{y}_{sdICL}$ to optimize $\phi$ while freezing $\theta$ as:
\begin{equation}
    \mathcal{L}_{D} = \max_{\phi} log~P(\hat{y}_{sdICL}|z_{sdICL};\theta;\phi)
\end{equation}
This objective should force the guarded LLM to produce identical probability given target data in ICL format for every label from the label set that the owner knows.

\mypara{Maintenance Loss}
Prior studies found that fine-tuning models for a specific downstream task can adversely affect their performance on other tasks~\cite{MGTD23}.
Our results in~\autoref{sec:loss_study} also align with these findings.
If we only fine-tune the model with $\mathcal{L}_{D}$, the ICL and regular LM function of the guarded LLM degenerate compared to the original LLM.
Therefore, we present the maintenance loss, $\mathcal{L}_{M}$, which aims to retain the ICL ability on non-target data using $\mathcal{D}_{sgICL}$.
Similar to how $\mathcal{L}_{D}$ operates, we also maximize the log-likelihood of $y_{sgICL}$ to optimize $\phi$ while $\theta$ remains fixed.
The loss $\mathcal{L}_{M}$ is computed as:
\begin{equation}
    \mathcal{L}_{M} = \max_{\phi} log~P(y_{sgICL}|z_{sgICL};\theta;\phi)
\end{equation}
This objective should compel the guarded LLM to produce a posterior similar to the original LLM on non-target data in ICL format, thereby maintaining the ICL performance.

\mypara{Utility Loss}
Both $\mathcal{L}_{D}$ and $\mathcal{L}_{M}$ regulate the model's ICL behavior on target and non-target data, respectively.
Next, we focus on the model utility of processing regular data, i.e., non-ICL format prompt.
Similar to the ICL performance drops on non-target data, we also observe that the performance of processing regular sentences on the guarded LLM is compromised.
A possible explanation for this might be the issue of ``concept forgetting''~\cite{MGTD23}.
Therefore, we introduce the utility loss $\mathcal{L}_{U}$ to re-align the guarded LLM with respect to the original LLM.
A straightforward approach is to fine-tune the model with $\mathcal{D}_{sg}$ using the autoregressive loss.
However, the model might overfit the dataset since the size of the surrogate dataset is much smaller than the pre-training dataset.
As an alternative, we minimize the distance between text representations from the guarded and original models, which was previously demonstrated to be effective in~\cite{MGTD23, ZIESW18}.
The token representation at $i$-th location from layer $j$ is extracted from the guarded and original LLMs, denoted as ${h'}^j_i$ and $h^j_i$, respectively.
In addition, extending the optimization scope across layers enhances the model utility performance~\cite{ZIESW18}.
Therefore, we extract all token representations from every $l$ layer, normalize each vector using the $L2$-norm as $\frac{h_i^j}{||h_i^j||_{2}}$, and concatenate them into a single vector.
The objective is to minimize the L2 distance between these sets of representations, and thus, the loss $\mathcal{L}_{A}$ is defined as:
\begin{equation}
    \mathcal{L}_{U} = \min_{\phi} ||h' - h||_{2}
\end{equation}
This objective should force the guarded LLM to produce similar outputs as the original LLM on all data.

\begin{algorithm}
\caption{\system}
\label{alg:algorithm}
\SetKwInput{KwInput}{Input}                
\SetKwInput{KwOutput}{Output}              
\Input{$\mathcal{D}_{sdICL}$, $\mathcal{D}_{sgICL}$, $\mathcal{D}_{sg}$, LLM $\theta$, PEFT module $\phi$}
\Output{PEFT module $\phi$}
\For{$t \leftarrow 0$ \KwTo $T$}{
    $(z_{sdICL}, \hat{y}_{sdICL}) \sim \mathcal{D}_{sdICL}$\;
    $\mathcal{L}_{D} = log~P(\hat{y}_{sdICL}|z_{sdICL};\theta;\phi)$\;
    
    $(z_{sgICL}, y_{sgICL}) \sim \mathcal{D}_{sgICL}$\;
    $\mathcal{L}_{M} = log~P(y_{sgICL}|z_{sgICL};\theta;\phi)$\;

    $x_{sg} \sim \mathcal{D}_{sg}$\;
    $\{{h'}_1, {h'}_2, \cdots, {h'}_n\} = \phi(\theta(x_{sg}))$\;
    $h' = ({h'}_1; {h'}_2; \cdots; {h'}_n)$\;
    $\{h_1, h_2, \cdots, h_n\} = \theta(x_{sg})$\;
    $h = (h_1; h_2; \cdots; h_n)$\;
    $\mathcal{L}_{U} = ||h' - h||_{2}$\;

    $\mathcal{L} = \mathcal{L}_{D} + \mathcal{L}_{M} + \mathcal{L}_{U}$\;
    updating $\phi$ with $\mathcal{L}$;\
}
\end{algorithm}

\subsection{Overall}

After presenting the building blocks, we introduce the overall process of \system.
First, the owner aims to guard the original LLM by fine-tuning the PEFT module and plugging it into the LLM.
For each epoch, \system samples the data from $\mathcal{D}_{sdICL}$ and obtains the output from the guarded LLM to compute the cross-entropy for $\mathcal{L}_{D}$.
Second, it computes the cross-entropy for $\mathcal{L}_{M}$ with the data from $\mathcal{D}_{sgICL}$.
Last, \system samples the data from $\mathcal{D}_{sg}$ to obtain a sequence of normized vector $\{{h'}_1, {h'}_2, \cdots, {h'}_n\}$ from the guarded LLM and $\{h_1, h_2, \cdots, h_n\}$ from the original LLM, where $n$ is the sentence length.
Then, \system concatenates them and computes $\mathcal{L}_{U}$ by minimizing the L2 distance between $h'$ and $h$.
The final objective function is:
\begin{equation}
\mathcal{L} = \mathcal{L}_{D} + \mathcal{L}_{M} + \mathcal{L}_{U}
\end{equation}
Importantly, \system only updates the PEFT module $\phi$, and the original LLM $\theta$ is frozen.
The entire optimization process is repeated for $T$ epochs. 
Further details can be found in~\autoref{alg:algorithm}.

\section{Experimental Setup}
\label{sec:exp_setup}

In this section, we describe our evaluation setup, including the datasets, implementation details, and evaluation metrics.

\subsection{Dataset}

We use the following five datasets to conduct our experiments.
\begin{itemize}
    \item \textbf{FP}~\cite{MSKWT14} contains sentences from financial news annotated as positive, negative, or neutral. 
    \item \textbf{SST-2}~\cite{WSMHLB19} includes sentences extracted from movie reviews. 
    Each sentence is annotated by humans to indicate its sentiment, either positive or negative.
    This dataset is particularly selected as it has a similar label set as FP, which allows us to examine the impact of \system when deactivating the ICL ability on FP but not SST-2.
    \item \textbf{TREC}~\cite{LR02} is a question classification dataset with six coarse class labels. 
    The labels include abbreviation, entity, description, human, location, and number.
    \item \textbf{AGnews}~\cite{ZZL15} contains news articles related to the world, sports, business, and science \& technology. 
    It is a topic classification dataset with respect to four classes.
    \item \textbf{Lambada}~\cite{PKLPBPBBF16} is a collection of narrative texts designed to evaluate the capabilities of LM in understanding texts. 
\end{itemize}
Each FP, SST-2, TREC, and Agnews are treated interchangeably as the target data, while the remaining is treated as the auxiliary (non-target) data.
Also, we sample 100 examples from each dataset for the evaluation.

\begin{table*}[!t]
\centering
\begin{tabular}{lccccc}
\toprule
\multicolumn{1}{l}{} & \multicolumn{1}{c}{Target Dataset} & \multicolumn{4}{c}{Auxiliary Datasets} \\
\midrule
Target  & Accuracy (\%) & FP (\%) & SST-2 (\%) & TREC (\%) & AGnews (\%)  \\
\midrule
FP     &  0.31 (-0.49) & -            & 0.80 (-0.13) & 0.68 (-0.02) & 0.82 (-0.08) \\
SST-2  &  0.41 (-0.52) & 0.71 (-0.09) & -            & 0.68 (-0.02) & 0.88 (-0.02) \\
TREC   &  0.13 (-0.57) & 0.76 (-0.04) & 0.93 (+0.00) & -            & 0.89 (-0.01) \\
AGnews &  0.21 (-0.69) & 0.62 (-0.18) & 0.92 (-0.01) & 0.71 (-0.01) & -            \\
\bottomrule
\end{tabular}
\caption{The ICL performance and changes of the guarded LLM.}
\label{table:main_acc}
\end{table*}

\begin{table*}[!t]
\centering
\begin{tabular}{lcccccc}
\toprule
\multicolumn{1}{l}{} & \multicolumn{1}{c}{Target Dataset} & \multicolumn{5}{c}{Auxiliary Datasets} \\
\midrule
Target  & $\Delta$ PPL   & FP & SST-2 & TREC & AGnews & Lambada  \\
\midrule
FP     &  +0.020 & -      & -2.164 & +0.024 & -0.004 & -0.012 \\
SST-2  &  +2.885 & +0.091 & -      & +0.172 & +0.006 & 0.012 \\
TREC   &  +0.044 & -0.004 & -2.190 & -      & -0.007 & +0.002 \\
AGnews &  +0.003 & +0.025 & -2.006 & +0.131 & -      & +0.006 \\
\bottomrule
\end{tabular}
\caption{The utility changes of the guarded LLM.}
\label{table:main_ppl}
\end{table*}

\subsection{Implementation Details}

All experiments are implemented using PyTorch~\cite{PyTorch} and Transformers~\cite{WDSCDMCRLFB19}. 
We describe the experimental setup used for evaluation.

\mypara{Models}
We experiment with 3 models: LLaMA~\cite{TLIMLLRGHARJGL23}, OPT~\cite{ZRGACCDDLLMOSSSKSWZ22}, and Cerebras~\cite{DGCKMPTH23}.
Each model is pre-trained on distinct datasets. 
Both OPT and Cerebras support English, whereas LLaMA can process multiple languages.
LLaMA has a vocabulary size of 32,000, which is smaller than the 50,257 of both OPT and Cerebras.
Besides, LLaMA (13B) outperforms OPT (13B) and Cerebras (13B) in several evaluations~\footnote{https://huggingface.co/spaces/HuggingFaceH4/open\_llm\_leaderboard}, including text reasoning, understanding, and generation.
It also demonstrates impressive zero-shot and few-shot capability.
Therefore, we primarily focus on LLaMA-13B.
We also assess a range of LLaMA model sizes, including 7B and 30B.

\mypara{Training}
To regulate ICL behavior, we apply the PEFT method on top of the original LLM.
We primarily experiment with LoRA on top of the LLaMA-13B through the paper unless specifically stated.
For LoRA, the rank is set at 8, alpha at 32, and the dropout rate is 0.1. 
In addition, we also evaluate \system with prompt tuning in the PEFT study.
For prompt tuning, we set the embedding length to 8 and 16, and the hidden size is the same as the LLM input embedding.
We use a batch size of 4 and a learning rate of 1 x $10^{-4}$ with the Adam optimizer.
We train the model for 20 epochs.~\footnote{We also explored training longer, but it did not improve performance.}

\mypara{Test Set}
Despite the success of ICL, the performance is unstable based on the choice of template and combination of demonstrations.
While many prior studies have explored the factors and potential mitigation approaches, the study is beyond the scope of this paper.
For simplicity, we utilize the minimal template as shown in~\autoref{app:template}.
To construct the test set, we random sample data from the training set to construct the demonstration with distinct random seeds from 40 to 50.
Then, we select the demonstration (random seed) that yields the highest performance for each dataset, and these sampled data are excluded from training.

\mypara{Other Details}
In all experiments presented in this paper, we default to using $k = 16$ input-label pairs and $u = 40$ unique input-label demonstrations unless stated otherwise. 
For the target data, 100 data points suffice to deactivate the ICL ability on target data. 
For the utility loss, we extract text representation from every $l = 2$ transformer layer.

\subsection{Evaluation Metrics}

We evaluate the performance of the guarded LLM using 2 metrics: ICL performance and utility.

\mypara{ICL Performance}
We evaluate the ICL performance of the guarded LLM using both the target and auxiliary datasets. 
For the target dataset, we compute the test set accuracy.
When the accuracy approximates random guessing, it indicates the effectiveness of deactivating the ICL ability on target data.
For the auxiliary dataset, we use datasets distinct from the target dataset to serve as auxiliary datasets and compute the accuracy of the auxiliary test set. 
If the accuracy closely aligns with the ICL performance of the original LLM, it demonstrates the effectiveness of persevering the ICL ability on non-target data.

\mypara{Utility}
Besides considering the ICL performance on target and non-target data, it is necessary to evaluate the model's utility.
We use perplexity (PPL) to evaluate the utility of LM:
\begin{equation*}
    PPL = exp\{ -\frac{1}{n} \sum_{i=1}^n log (w_i|w_{1 < n}) \}
\end{equation*}
PPL is a metric to measure the quality of sentence generation, i.e., how well the model predicts $w_{n+1}$ given the proceeding sequence $w_{1<n}$.
While a lower perplexity typically indicates greater utility, we care about the PPL change between the guarded and original LLM in this evaluation.
The utility of the guarded LLM should align with the original LLM, given all data is in non-ICL format.
Hence, we compute the PPL change for regular sentences between the two models:
\begin{equation*}
    \Delta PPL = PPL_{original} - PPL_{guarded}
\end{equation*}

\begin{figure*}[!t]
\centering
\begin{subfigure}[!t]{0.5\columnwidth}
\centering
\includegraphics[width=\columnwidth]{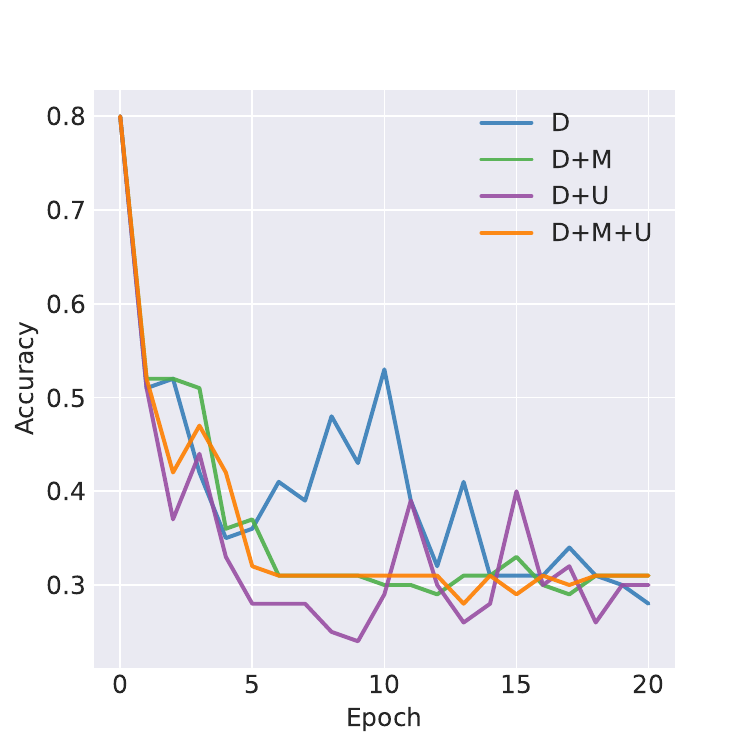}
\caption{FP}
\label{fig:fp_acc}
\end{subfigure}
\begin{subfigure}[!t]{0.5\columnwidth}
\centering
\includegraphics[width=\columnwidth]{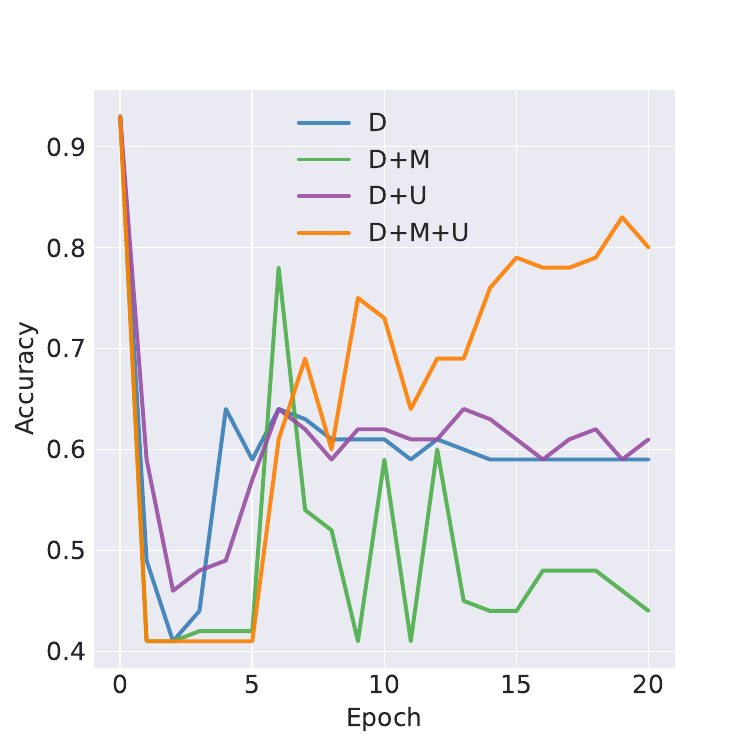}
\caption{SST-2}
\label{fig:sst2_acc}
\end{subfigure}
\begin{subfigure}[!t]{0.5\columnwidth}
\centering
\includegraphics[width=\columnwidth]{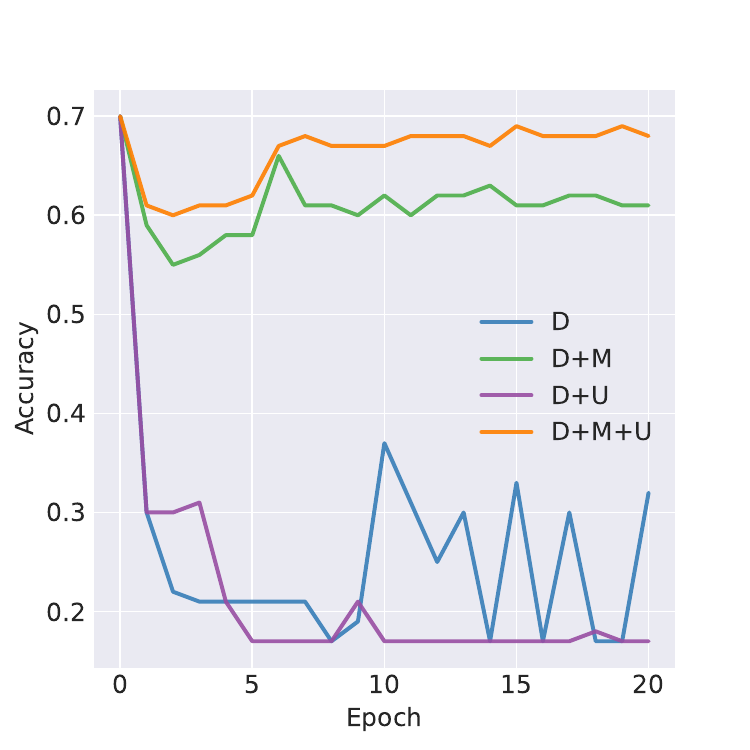}
\caption{TREC}
\label{fig:trec_acc}
\end{subfigure}
\begin{subfigure}[!t]{0.5\columnwidth}
\centering
\includegraphics[width=\columnwidth]{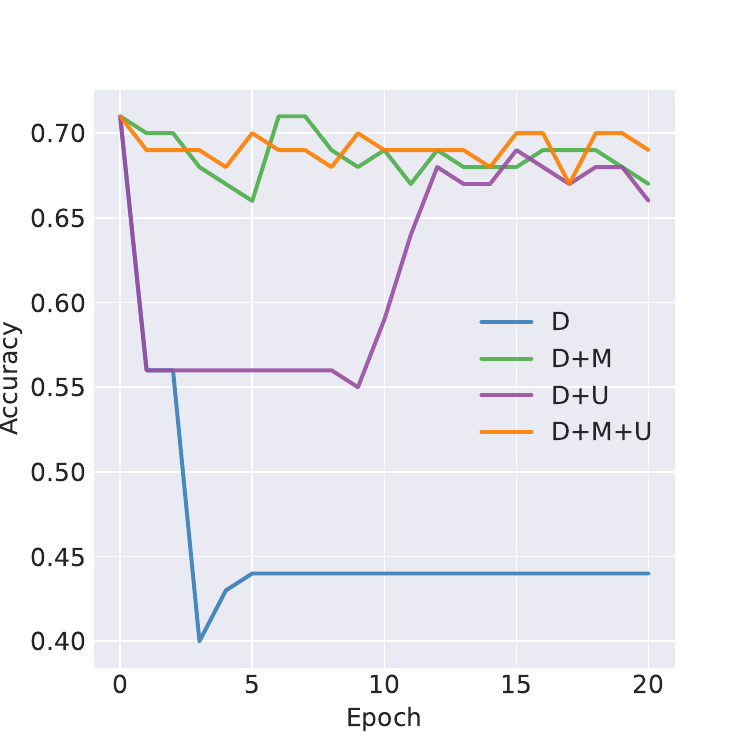}
\caption{AGnews}
\label{fig:agnews_acc}
\end{subfigure}
\caption{The ICL performance changes during fine-tuning under different loss combinations.
D = disable loss, M = maintenance loss, U = utility loss.}
\label{fig:loss_study}
\end{figure*}

\section{Evaluation}
\label{sec:evaluation}

We examine the experimental results from different perspectives. 
First, we evaluate the overall performance of the guarded LLM while deactivating the ICL ability on different target data.
Second, we investigate the effectiveness of each loss function and the impact. 
Following this, we explore different generation strategies for constructing the surrogate dataset and examine the potential for bypassing the ICL applicability authorization.
Finally, we study the effects of varying setups on \system.
We provide the original ICL performance and utility of LLaMA-13B for each dataset in~\autoref{table:baseline}.

\subsection{General} 
\label{sec:general}

We first evaluate the ICL performance of the guarded LLM after deactivating the ICL ability on target data (i.e., FP, SST-2, TREC, and AGnews).
~\autoref{table:main_acc} illustrates the accuracy and changes compared to the original LLM.
In general, the guarded LLM can deactivate the ICL ability on target data and bring the accuracy close to random guessing.
Meanwhile, the ICL performance on auxiliary datasets only drops negligibly overall.
For instance, when the target dataset is TREC, its accuracy drops from 70\% to 13\%, while the accuracy on FP, SST-2, and AGnews is maintained at 76\%, 93\%, and 89\%, respectively.
However, we observe that the ICL performance on the auxiliary dataset is affected in some cases.
When the target object is FP, the accuracy on SST-2 drops by 13\%, which is slightly higher than other auxiliary datasets.
This unexpected result could be due to the overlapping label set between FP and SST-2, and deactivating the ICL ability on FP might also affect SST-2 accidentally. 
We observe the same phenomenon when targeting SST-2 as the accuracy on FP drops by 9\%.

Next, we evaluate the utility of the guarded LLM.
As shown in~\autoref{table:main_ppl}, the PPL changes are negligible across target and auxiliary datasets.
The result demonstrates that the guarded LLM can produce outputs almost identical to the original on regular prompts.
Note that the PPL change on SST-2 is generally higher since it has a higher PPL on the original LLM.
Overall, these results indicate the efficacy of \system to fine-tune the guarded LLM on deactivating the ICL ability on target data without compromising the ICL performance on non-target data and the regular LM function on all data simultaneously.
In the following section, we primarily use FP as the target dataset to investigate the effectiveness of \system from other aspects.

\subsection{Impact of the Loss Function}
\label{sec:loss_study}

In this section, we investigate the impact of each loss function on controlling the ICL ability and regular LM ability on target and auxiliary data.
In~\autoref{table:loss_acc} and \autoref{table:loss_ppl}, we first note that using the disable loss alone can deactivate the ICL ability on FP completely from 80\% to 28\%.
As expected, the ICL ability on auxiliary datasets is affected significantly.
For instance, SST-2, TREC, and AGnews have a 34\%, 38\%, and 67\% accuracy drop, respectively.
The PPL change on them is also massive, with more than 20.
All these results reflect that the guarded LLM has altered completely.

\begin{table}[!t]
\centering
\resizebox{\columnwidth}{!}{
\begin{tabular}{lcccc}
\toprule
 & Target & \multicolumn{3}{c}{Auxiliary} \\
\midrule
Loss & FP (\%) & SST-2 (\%) & TREC (\%) & AGnews (\%) \\
\midrule
D                   & -0.52 & -0.34 & -0.38 & -0.67 \\
D+M                 & -0.49 & -0.49 & -0.09 & -0.29 \\
D+U                 & -0.50 & -0.32 & -0.53 & -0.70 \\
D+M+U               & -0.49 & -0.13 & -0.02 & -0.08 \\ 
\bottomrule
\end{tabular}
}
\caption{The ICL performance changes with different loss combinations.
D = disable loss, M = maintenance loss, U = utility loss.}
\label{table:loss_acc}
\end{table}

\begin{table}[!t]
\centering
\resizebox{\columnwidth}{!}{
\begin{tabular}{lccccc}
\toprule
 & Target & \multicolumn{4}{c}{Auxiliary} \\
\midrule
Loss & FP & SST-2 & TREC & AGnews & Lambada \\
\midrule
D                                 & +24.08 & -198.5  & +20.75 & +48.89 & +32.12 \\
D+M                & -0.484  & -18.46   & -0.192  & -0.062  & -0.099  \\
D+U                 & +0.004  & -2.220    & -0.039  & -0.007  & -0.006  \\
D+M+U & +0.020  & -2.164    & +0.024  & -0.004  & -0.012  \\ 
\bottomrule
\end{tabular}
}
\caption{The utility changes with different loss combinations.
D = disable loss, M = maintenance loss, U = utility loss.}
\label{table:loss_ppl}
\end{table}

Then, when we optimize the model with both disable and maintenance loss, the ICL ability on auxiliary datasets recovers to a certain extent without sacrificing the deactivating performance.
For instance, TREC and AGnews have only 9\% and 29\% accuracy drops now, and the PPL changes are also much lower.
Although the overall performance is not the best, the guarded LLM can function adequately close to the original LLM on data in non-ICL format. 
Besides, we note that both the ICL performance and utility on SST-2 are affected significantly.
As mentioned in~\autoref{sec:general}, the result is likely to be related to the overlapping label set.
Therefore, fine-tuning the model with both disable and maintenance loss together might hurt the optimization. 

On the other hand, we find that the utility changes across datasets are negligible when using both disable and utility loss.
It confirms that the utility loss effectively preserves the guarded LLM's functionality in processing regular data.
Interestingly, we observe that the ICL performance on SST-2 is better than using the maintenance loss as the accuracy goes up to 61\%.
Taken together, using all three losses produces the best performance in controlling ICL behavior and maintaining the LLM functionality.

\begin{table}[!t]
\centering
\resizebox{\columnwidth}{!}{
\begin{tabular}{lcccc}
\toprule
 & Target & \multicolumn{3}{c}{Auxiliary} \\
\midrule
Method & FP (\%) & SST-2 (\%) & TREC (\%) & AGnews (\%) \\
\midrule
Random         & -0.49 & -0.13   & -0.02 & -0.08 \\ 
Auxiliary (SST-2)             & -0.62 & +0.00   & -0.02 & -0.06 \\
JS             & -0.50 & -0.12   & +0.00 & +0.00 \\
\midrule
M-Cos        & -0.66 & -0.10   & -0.01 & +0.00 \\
M-L2         & -0.65 & -0.09   & -0.01 & +0.00 \\
\midrule
L-Cos        & -0.49 & -0.50   & +0.00 & +0.00 \\
L-L2         & -0.63 & -0.24   & -0.06 & -0.04 \\
\bottomrule
\end{tabular}
}
\caption{The ICL performance changes with different generation strategies.
JS = Jaccard Similarity, FP=, M = MiniLM, L = LLaMA, Cos = cosine similarity, L2 = Euclidean distance.}
\label{table:sample_acc}
\end{table}

\begin{table}[!t]
\centering
\resizebox{\columnwidth}{!}{
\begin{tabular}{lccccc}
\toprule
 & Target & \multicolumn{4}{c}{Auxiliary} \\
\midrule
Method & FP & SST-2 & TREC & AGnews & Lambada \\
\midrule
Random         & +0.020 & -2.164 & +0.024 & -0.004 & -0.012 \\ 
Auxiliary (SST-2)             & +0.000 & +0.332 & +0.105 & +0.000 & +0.019 \\
JS             & +0.022 & +3.085 & +0.005 & -0.009 & -0.001 \\
\midrule
M-Cos        & -0.005 & +2.814 & +0.178 & +0.002 & -0.001 \\
M-L2         & +0.011 & +0.468 & +0.244 & +0.001 & -0.008 \\
\midrule
L-Cos        & +0.022 & +3.187 & +0.009 & +0.002 & -0.001 \\
L-L2         & +0.004 & +3.058 & +0.028 & +0.002 & +0.012 \\
\bottomrule
\end{tabular}
}
\caption{The utility changes with different generation strategies.
JS = Jaccard Similarity, M = MiniLM, L = LLaMA, Cos = cosine similarity, L2 = Euclidean distance.}
\label{table:sample_ppl}
\end{table}

\mypara{Fine-Tuning Process}
Furthermore, we report the ICL performance changes over the fine-tuning process in~\autoref{fig:loss_study}.
First, with no surprise, all loss combinations can deactivate the ICL ability on FP in the early stage, as shown in~\autoref{fig:fp_acc}.
Second, we observe that the ICL ability on the auxiliary dataset is affected at the beginning of the fine-tuning, but it is recovered later.
For instance, the accuracy on TREC drops initially and is recovered after 5 epochs, as depicted in~\autoref{fig:trec_acc}.
Besides, we observe that the performance on SST-2 fluctuates when using only disable and maintenance loss during the fine-tuning in~\autoref{fig:sst2_acc}.
It confirms our hypothesis that the disable and maintenance loss conflict with each other, leading to unstable optimization.

\mypara{Takeaways}
In summary, the disable loss can deactivate the ICL ability on target data, the maintenance loss retains the ICL ability on non-target data, and the utility loss ensures the regular function of the LLM on all data. 
Implementing all three losses simultaneously leads to successful ICL applicability authorization on the LLM.

\begin{figure}[!t]
\centering
\begin{subfigure}[!t]{0.45\columnwidth}
\centering
\includegraphics[width=\columnwidth]{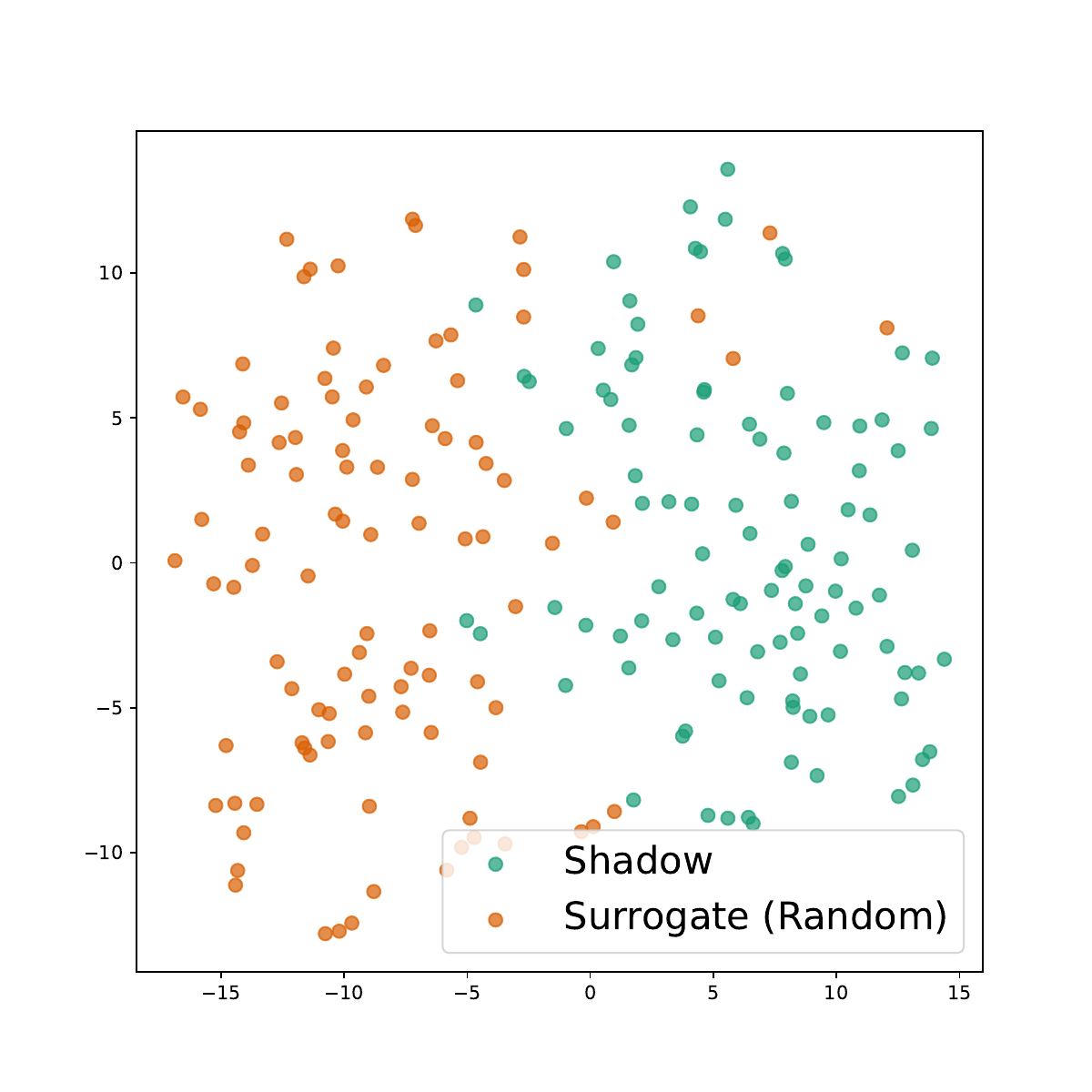}
\caption{Random}
\label{fig:minlm_old}
\end{subfigure}
\begin{subfigure}[!t]{0.45\columnwidth}
\centering
\includegraphics[width=\columnwidth]{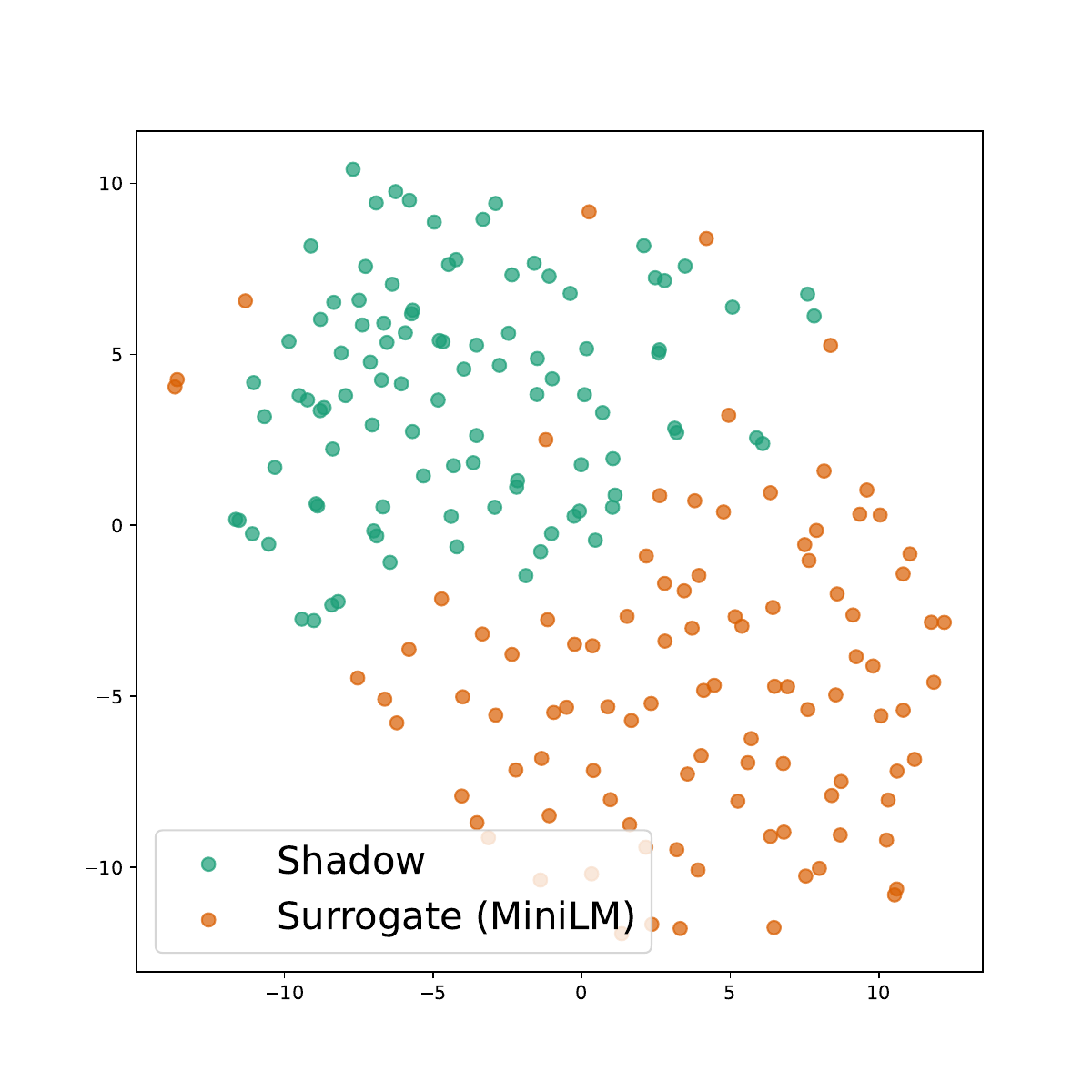}
\caption{MiniLM-L2}
\label{fig:minlm_new}
\end{subfigure}
\caption{The t-SNE visualization on the shadow and surrogate data with MiniLM using L2 distance.}
\label{fig:minlm_tsne}
\end{figure}

\begin{figure}[!t]
\centering
\begin{subfigure}[!t]{0.45\columnwidth}
\centering
\includegraphics[width=\columnwidth]{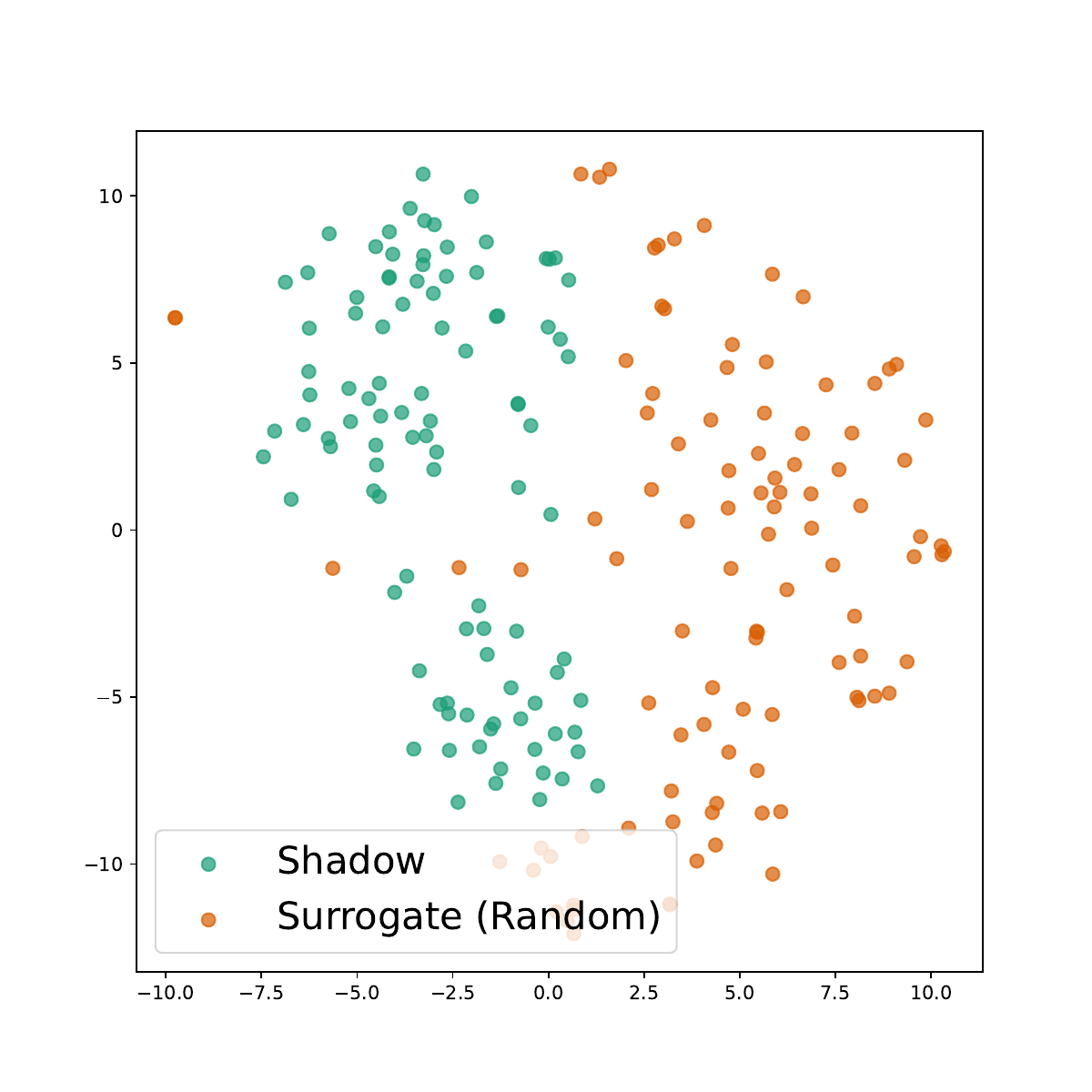}
\caption{Random}
\label{fig:llama_old}
\end{subfigure}
\begin{subfigure}[!t]{0.45\columnwidth}
\centering
\includegraphics[width=\columnwidth]{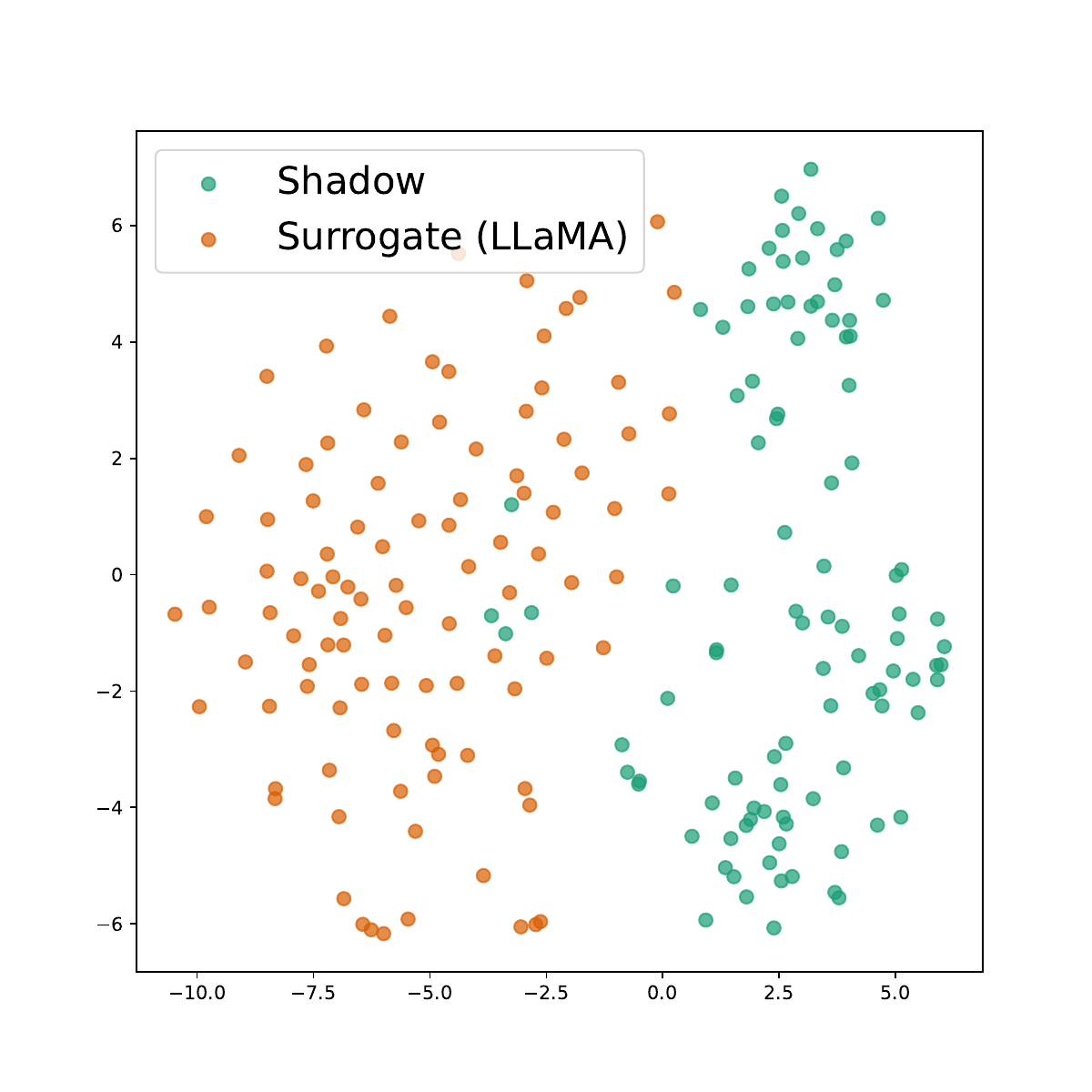}
\caption{LLaMA-L2}
\label{fig:llama_new}
\end{subfigure}
\caption{The t-SNE visualization on the shadow and surrogate data with LLaMA-13B using L2 distance.}
\label{fig:llama_tsne}
\end{figure}

\begin{figure*}[!t]
\centering
\begin{subfigure}[!t]{0.65\columnwidth}
\centering
\includegraphics[width=\columnwidth]{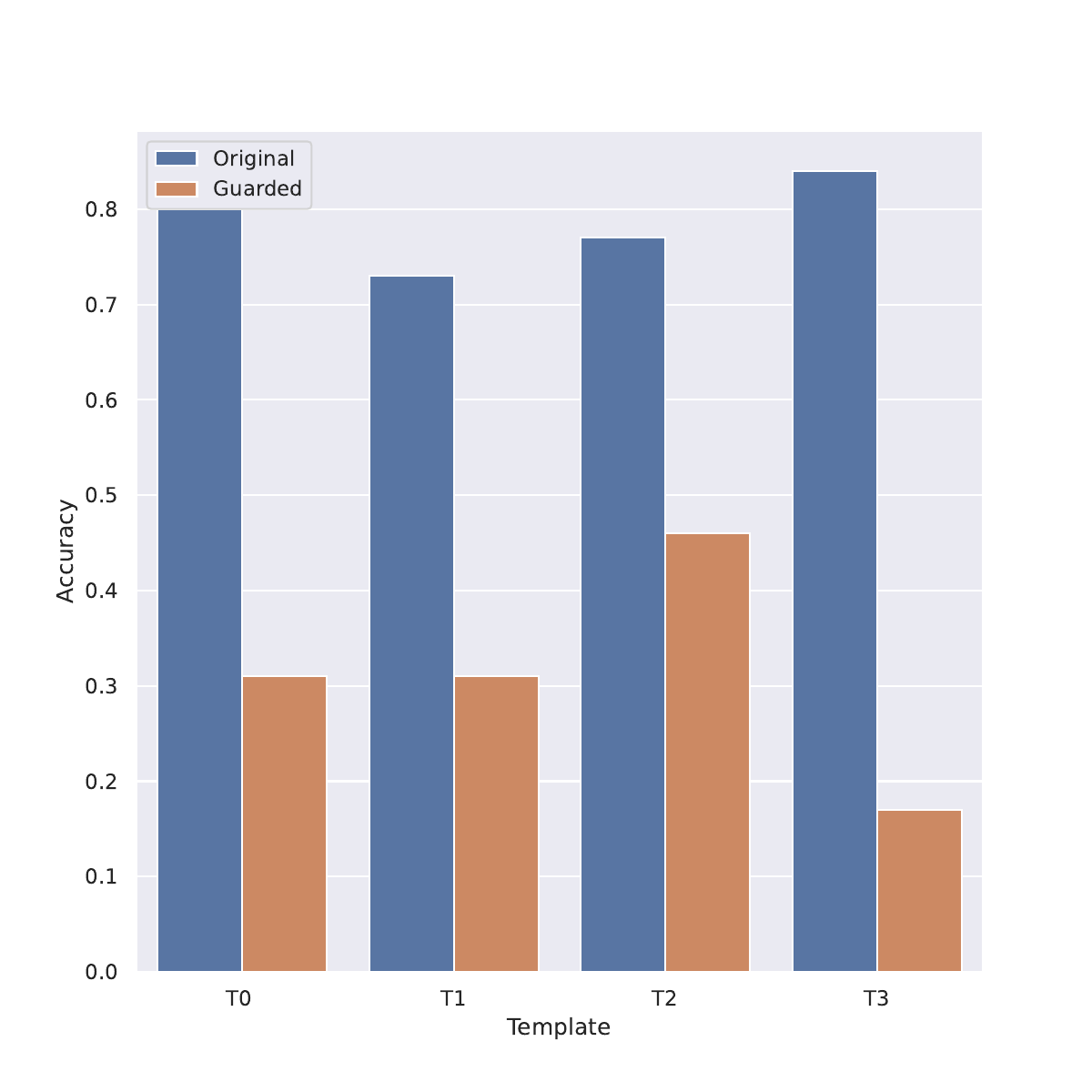}
\caption{Different templates.}
\label{fig:temp_exp}
\end{subfigure}
\begin{subfigure}[!t]{0.65\columnwidth}
\centering
\includegraphics[width=\columnwidth]{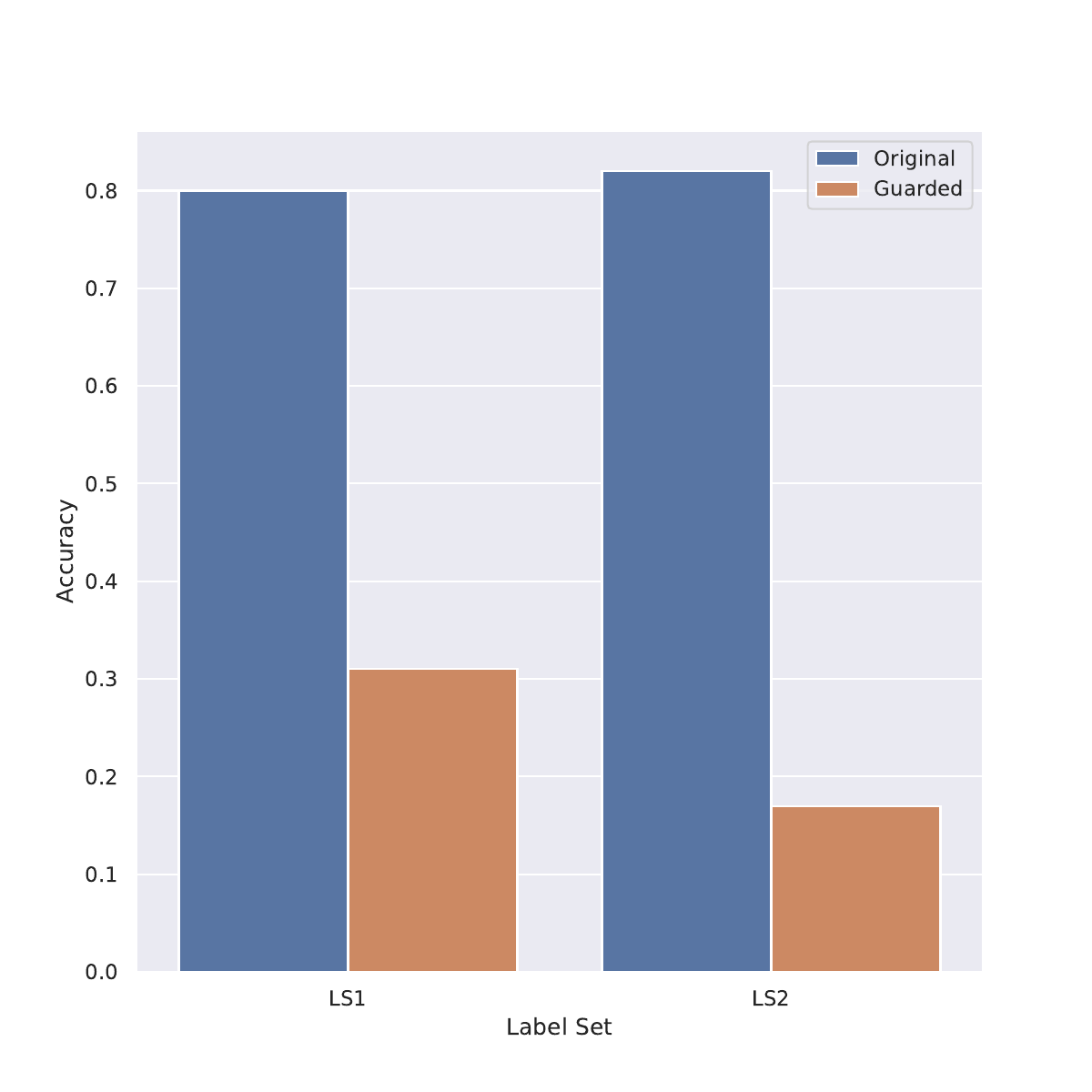}
\caption{Different label sets.}
\label{fig:label_exp}
\end{subfigure}
\begin{subfigure}[!t]{0.65\columnwidth}
\centering
\includegraphics[width=\columnwidth]{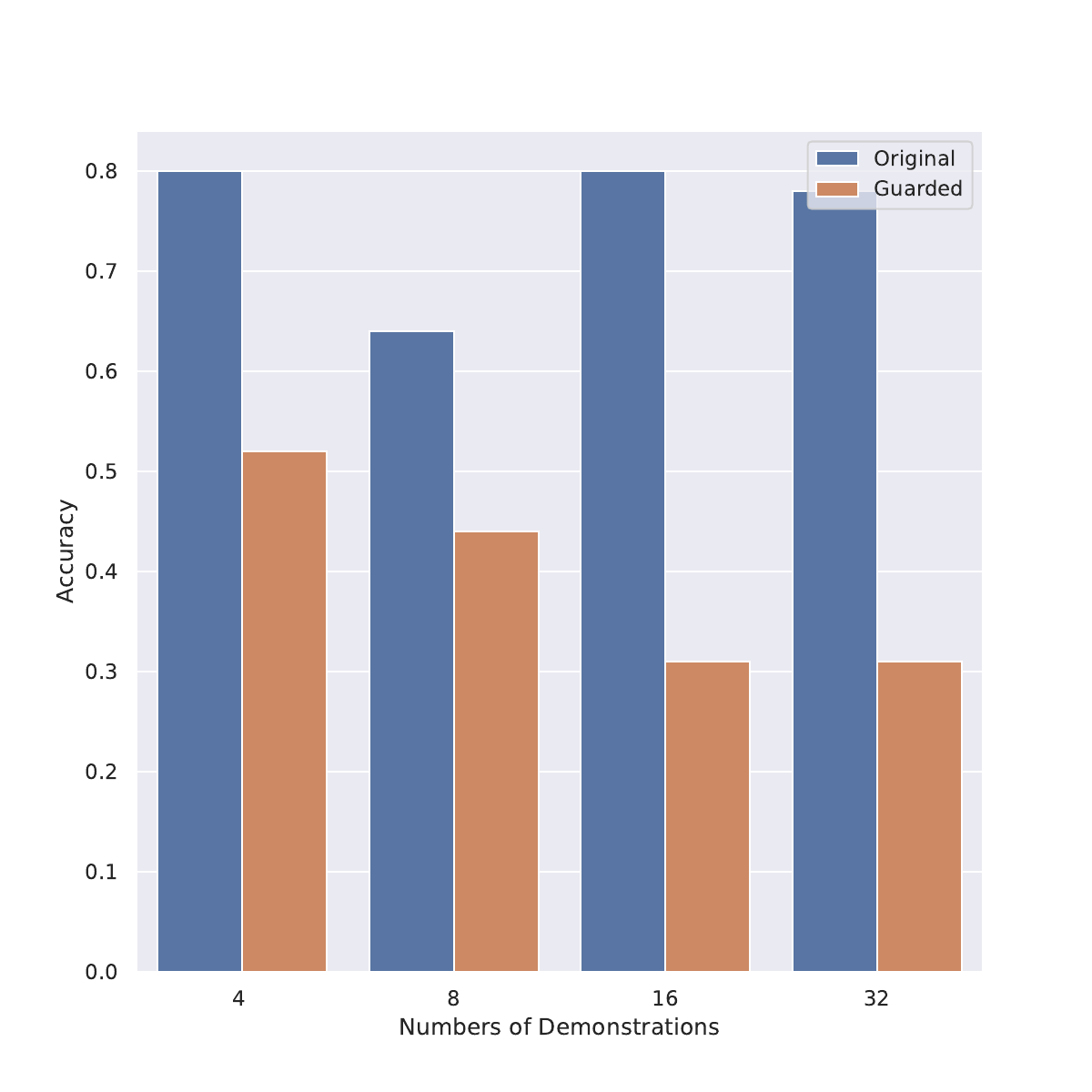}
\caption{Different numbers of demonstrations.}
\label{fig:demo_exp}
\end{subfigure}
\caption{The ICL performance on FP against the adaptive adversary.}
\label{fig:dome_bar}
\end{figure*}

\subsection{Surrogate Dataset Augmentation}
\label{sec:sample_study}

In~\autoref{sec:methodology}, we build the surrogate dataset by random sampling data from the original LLM.
However, the sampled data can be similar to the target data, which might hurt the optimization.
If the sampled data is similar to auxiliary data like SST-2, we might mitigate the ICL performance drops.
To justify our hypothesis, we first assume that we have access to the SST-2 dataset and use it to replace the surrogate dataset.
In~\autoref{table:sample_acc} and~\autoref{table:sample_ppl}, we observe that the ICL performance of the guarded LLM is now identical to the original LLM as the accuracy change on SST-2 is 0\%.
The results demonstrate that a good surrogate dataset can reduce the impact of fine-tuning.
Since the auxiliary or non-target data is unavailable, our best solution is to generate sample data that differs from the target and other sample data.
Concretely, we aim to produce data by increasing the sentence distance between the shadow and surrogate datasets.
We consider the distance in terms of discrete token and latent representation space.

\mypara{Discrete Token}
First, we attempt to generate data that is different with respect to the token level.
We use Jaccard Similarity (JS) to measure the distance between target and sample data in terms of token.
JS computes the overlapping ratio between two text sentences:
\begin{equation}
    JS = \frac{x_1 \bigcap x_2}{x_1 \bigcup x_2}
\end{equation}
It evaluates how closely the two sentences are contextually related by calculating the proportion of common tokens to total tokens.
The idea here is to generate data with tokens different from the target and other sample data.
In detail, we retain only the data with a JS score of less than 0.1, meaning a maximum of 10\% token overlap between the target and sample data.
In~\autoref{table:sample_acc}, we find that it can improve the ICL ability on TREC and AGnews, which has a 0\% accuracy drop now.
In addition, the utility remains decent as the PPL changes are still negligible across datasets.

\mypara{Latent Representation}
Second, we explore to generate data that is different in latent space.
We use the original LLaMA (conditional LM) and MiniLM~\footnote{https://www.sbert.net/index.html} (Masked LM) to extract the sentence representation.
For MiniLM, we obtain the sentence representation by applying the mean pooling on top of the token representation.
However, LLaMA is a conditional LM, where process input is from left to right.
Hence, the first token representation does not contain any information about the later token, which is not ideal for applying mean pooling.
Instead, we use the last token representation to approximate the sentence representation, which has attended to all the tokens in the sentence.
We then measure the sentence distance with the Euclidean distance (L2) or cosine similarity (Cos) and only keep the generated data when the score is larger or smaller than the threshold respectively.~\footnote{We determine the threshold by measuring the score between data points in the target dataset.}
Our objective is to produce data that covers more latent space than random sampling.

It can be seen from the data in~\autoref{table:sample_acc} that using the representation from MiniLM outperforms LLaMA in general.
For instance, generating data based on L2 or Cos with MiniLM decreases the ICL performance on FP further to 14\% and 15\% and mitigates the impact on TREC and AGnews.
Also, using L2 distance can preserve the ICL ability on SST-2 better up to 84\% accuracy.
On the other hand, using the representation from LLaMA does not improve the performance of controlling ICL behavior in general.
For instance, using Cos with LLaMA can also mitigate the impact on TREC and AGnews, but the ICL performance on SST-2 is worse than the random sampling.
While using L2 with LLaMA can decrease the ICL performance on FP to 17\%, the ICL performance on SST-2 also drops to 69\%.
In~\autoref{table:sample_ppl}, we report the PPL changes.
What stands out in the table is that using L2 with MiniLM produces better utility on SST-2 with a PPL change of 0.468, which is close to using FP as the surrogate data.
It shows the effectiveness of generating data with MiniLM based on L2 distance.
We also visualize the shadow and surrogate data in~\autoref{fig:minlm_tsne} and~\autoref{fig:llama_tsne}.
As the figure shows, generating data with L2 distance indeed has more space coverage than random sampling, and the newly generated data has less overlapping with the shadow data.

\mypara{Takeaways}
In summary, compared to random sampling, we find that generating data that is distant in token level or latent space produces better performance in terms of the ICL ability and utility of the guarded LLM.

\subsection{Adaptive Attacks}
\label{sec:adaptive_study}

In~\autoref{sec:methodology} and~\autoref{sec:exp_setup}, we present the details of constructing the shadow and surrogate ICL dataset and set up the demonstration (16 examples) using the minimal template with the default label set.
In practical scenarios, the user has the freedom to decide the use of the template, label set, and number of demonstrations.
If the malicious user realizes that some demonstration setups are ineffective, they can switch to another setup, and we consider it an adaptive attack.
In this section, we evaluate the robustness of the guarded LLM against these adaptive demonstrations by considering three variables: template, label, and number of demonstrations.

\mypara{Template}
First, we conduct our experiment by using the other three templates in~\autoref{app:template}. 
As depicted in~\autoref{fig:temp_exp}, the guarded LLM remains effective at deactivating the ICL ability on FP, even when encountering templates that are not presented during the fine-tuning phase. 
We observe that the ICL ability on FP with template 1 and 3 still has over 40\% and 60\% accuracy drops, respectively.
On the other hand, using template 2 can perverse the ICL performance slightly up to 46\%, but that is not enough to be usable for the adversary.

\mypara{Label}
Second, we investigate the implications of label mismatches using other label sets in~\autoref{app:template}.
Apart from altering the template, modifying the label is another viable avenue for adversaries.
In~\autoref{fig:label_exp}, we find that switching labels does not undermine the ICL applicability authorization.
Instead, the accuracy on FP drops further from 31\% to 17\%.

\mypara{Number of Demonstrations}
While our study constructs the shadow and surrogate ICL dataset utilizing 16 demonstrations, an adversary might employ a different quantity (e.g., 4, 8, or 32).
We conduct the experiment to investigate the impact of varying the number of demonstrations.
As shown in~\autoref{fig:demo_exp}, reducing the number of demonstrations marginally weakens the guarded LLM's ability to deactivate ICL on FP.
For instance, the accuracy increases from 31\% to 52\% as the number of demonstrations decreases from 16 to 4.
But still, the adversary should not be able to use ICL to perform the task with low accuracy.

\mypara{Takeaways}
In general, the guarded LLM is able to deal with ICL prompts with different setups.
Although we can see some drops in certain cases, we are not concerned about it since the ICL performance on target data is still maintained at a low level.

\subsection{Further Study}
\label{sec:further_study}

\begin{table}[!t]
\centering
\resizebox{\columnwidth}{!}{
\begin{tabular}{lcccc}
\toprule
 & Target & \multicolumn{3}{c}{Auxiliary} \\
\midrule
Models       & FP (\%) & SST-2 (\%)   & TREC (\%) & AGnews (\%) \\
\midrule
LLaMA-13B    & 0.31 (-0.49) & 0.80 (-0.13) & 0.68 (-0.02) & 0.82 (-0.08) \\
OPT-13B      & 0.17 (-0.56) & 0.86 (-0.07) & 0.59 (+0.03) & 0.30 (-0.48) \\
Cerebras-13B & 0.31 (-0.56) & 0.81 (-0.08) & 0.40 (-0.01) & 0.37 (-0.40) \\
\bottomrule
\end{tabular}
}
\caption{The ICL performance with different models.}
\label{table:model_acc}
\end{table}

\begin{table}[!t]
\centering
\resizebox{\columnwidth}{!}{
\begin{tabular}{lcccc}
\toprule
 & Target & \multicolumn{3}{c}{Auxiliary} \\
\midrule
Models       & FP (\%) & SST-2 (\%)   & TREC (\%) & AGnews (\%) \\
\midrule
7B     & 0.51 (-0.26) & 0.92 (-0.03) & 0.77 (-0.03) & 0.85 (-0.04) \\
13B    & 0.31 (-0.49) & 0.80 (-0.13) & 0.68 (-0.02) & 0.82 (-0.08) \\
30B    & 0.31 (-0.52) & 0.91 (-0.04) & 0.92 (+0.00) & 0.91 (+0.00) \\
\bottomrule
\end{tabular}
}
\caption{The ICL performance with different model sizes of LLaMA model.}
\label{table:modelsize_acc}
\end{table}

We now investigate the effects of different setups for \system. 
First, we examine the impact of applying \system to other models and the influence of model size. 
Second, we demonstrate the effectiveness of deactivating multiple datasets. 
Third, we compare performance based on the number of unique demonstrations and different amounts of sampling data. 
Finally, we evaluate the performance of \system compared to existing fine-tuning methods, and the performance between using LoRA and prompt tuning.

\mypara{Models}
We start by evaluating the efficacy of \system across different models: LLaMA-13B, OPT-13B, and Cerebras-13B.
The objective is to ascertain that \system can be applied universally across varied LLMs. 
We provide the original ICL performance and utility for OPT and Cerebras in~\autoref{app:baseline}.
In~\autoref{table:model_acc}, all guarded LLMs successfully reduce the ICL performance on FP to nearly random guessing. 
We observe that both OPT and Cerebras have better ICL performance (less performance drops) on SST-2 with 86\% and 81\% accuracy, respectively.
Besides, they exhibit a higher accuracy drop in using ICL on AGnews, achieving only 30\% and 37\% accuracy.
This result may be explained by the fact that AGnews data is similar to the pre-training data of these models.
In~\autoref{table:model_ppl} (\autoref{app:results}), AGnews generally has lower perplexity, and even mirror changes to the model with AGnews data would likely affect the model.
But still, the overall performance is comparable to the original model.

\begin{figure}[!t]
\centering
\includegraphics[width=0.8\columnwidth]{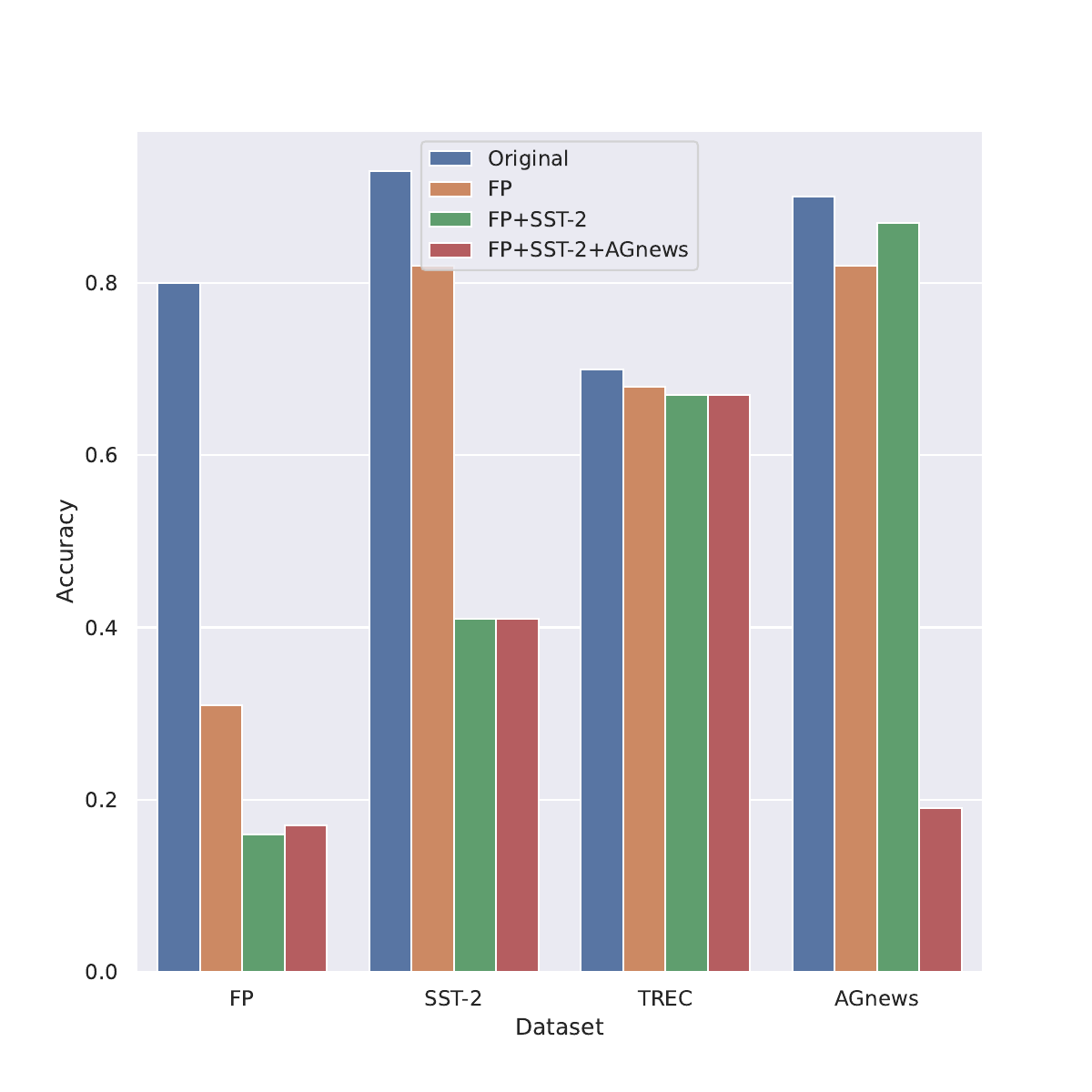}
\caption{The ICL performance of deactivating different multiple target datasets.}
\label{fig:multi_acc}
\end{figure}

\mypara{Model Size}
We then extended our evaluation by varying the model size of LLaMA from 7B to 30B, as shown in~\autoref{table:modelsize_acc}.
We provide the original ICL performance and utility for the 7B and 30B models in~\autoref{app:baseline}.
We observed that the smaller model has a weaker capability to deactivate the ICL ability on FP. 
In contrast, the 13B and 30B models can completely deactivate the ICL ability on FP.
Since the fine-tuning effect of \system on the 7B model is smaller, it also has a smaller impact on SST-2 and AGnews.
On the other hand, the 30B model is able to deactivate the ICL ability on FP without compromising the ICL performance on other auxiliary datasets, including SST-2.
\autoref{table:modelsize_ppl} (\autoref{app:results}) reports the utility changes, demonstrating that the impact of \system on the 7B and 30B models is less significant.

\begin{figure}[!t]
\centering
\includegraphics[width=.9\columnwidth]{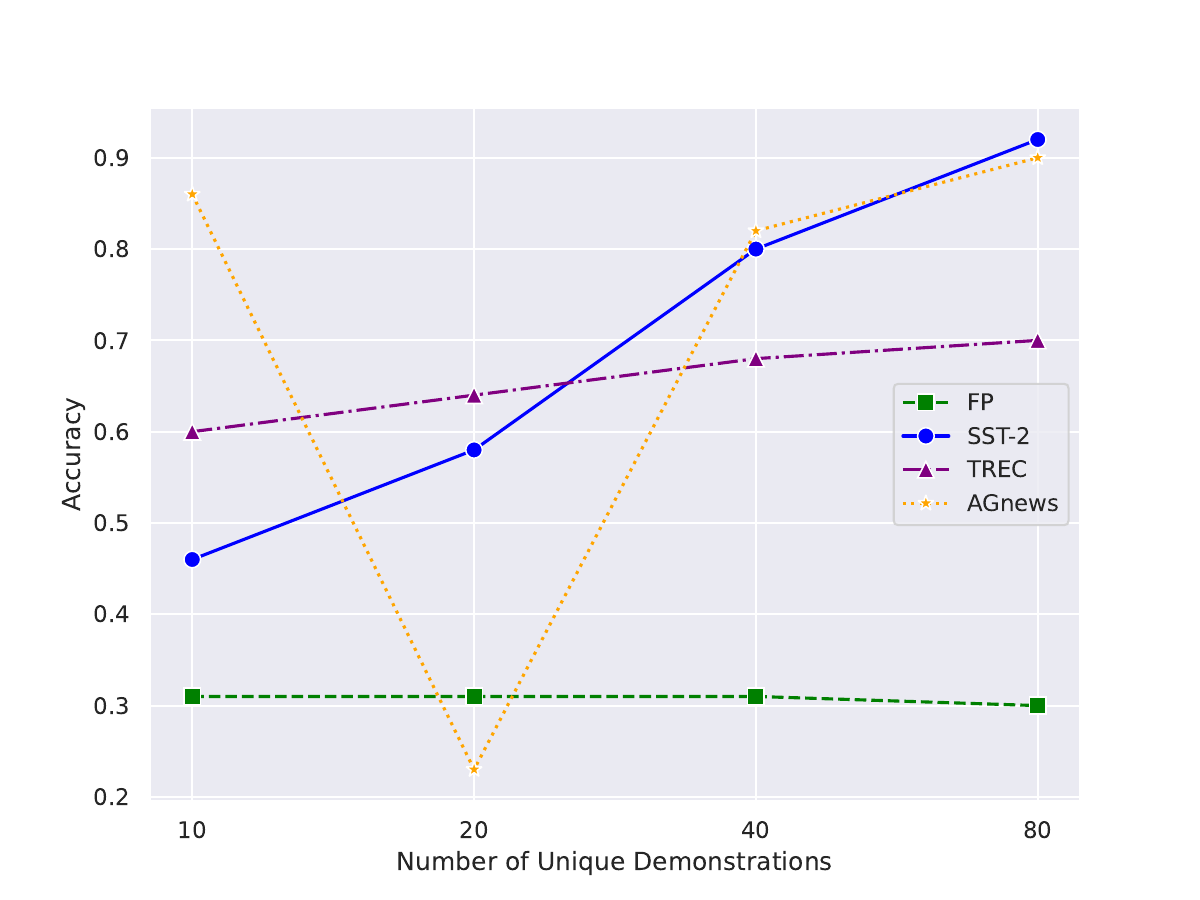}
\caption{The ICL performance with different numbers of unique demonstrations.}
\label{fig:unique_demo_exp}
\end{figure}

\mypara{Multiple Target}
In reality, the model owner might need to deactivate multiple data simultaneously.
Therefore, we study the effectiveness of \system in deactivating multiple datasets in two scenarios: 1) FP and SST-2, and 2) FP, SST-2, and AGnews. 
As shown in \autoref{fig:multi_acc} and \autoref{table:multi_ppl}, \system successfully deactivates multiple targets without compromising the model utility. 
In the first scenario, the accuracy of FP and SST-2 drops to 16\% and 41\%, respectively, while the accuracy of TREC and AGnews only decreases by 3\% and 3\%.
In the second scenario, the accuracy of FP, SST-2, and AGnews drops to 17\%, 41\%, and 19\%, respectively. 
Although there is a slight change in the accuracy and utility of TREC in both scenarios, the general ICL performance and model utility remain comparable to the original model. 
The results indicate the effectiveness of using \system to manage multiple ICL control requirements.

\mypara{Number of Unique Demonstrations}
We evaluate the effect of varying the number of unique demonstrations from 10 to 80. 
Typically, increasing the number of unique demonstrations should enhance the model's ability to control ICL behavior on both target and non-target data.
In~\autoref{fig:unique_demo_exp}, the results reveal that even with a smaller number of unique demonstrations, it is still possible to deactivate the ICL ability on FP.
However, we also suspect that performance drops at 20 demonstrations are due to the instability of ICL, and using a larger number of unique demonstrations can mitigate the issue.
In addition, using more unique demonstrations can preserve the ICL ability on other datasets better.
For instance, the accuracy on SST-2 increases from 46\% to 90\%, and for TREC, it increases from 60\% to 70\%.
These relationships may be explained by using more unique demonstrations, which enable the guarded LLM to better distinguish the data from SST-2 or FP during fine-tuning.
In~\autoref{table:uniquedemo_ppl} (\autoref{app:results}), we observe that the PPL change on TREC decreases as the number of unique demonstrations increases, although the difference is negligible.

\begin{figure}[!t]
\centering
\includegraphics[width=.9\columnwidth]{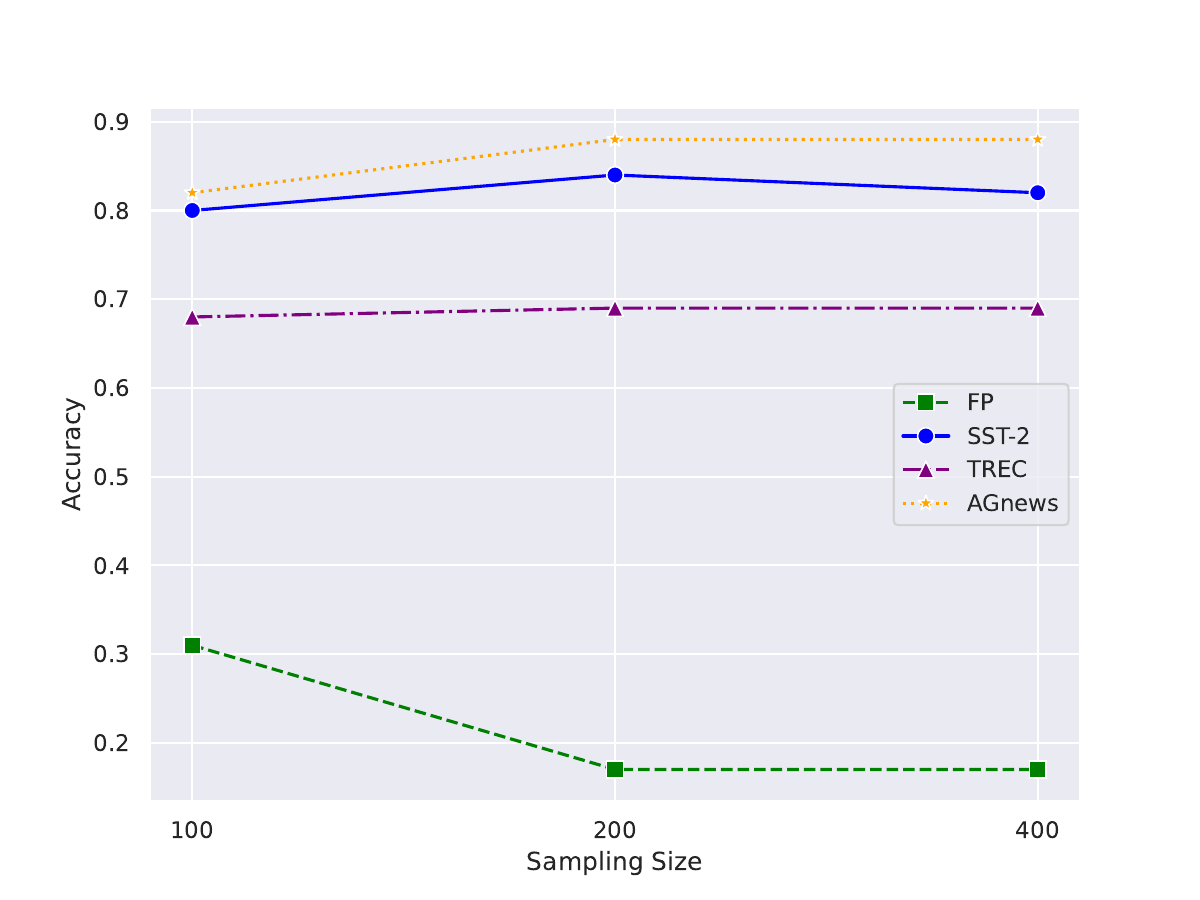}
\caption{The ICL performance with different sampling sizes.}
\label{fig:sample_size_acc}
\end{figure}

\mypara{Sampling Size}
In addition to selecting the number of unique demonstrations, the owner can also choose to sample more data from the original LLM.
We investigate the impact of varying sampling sizes (100, 200, 400) and believe this study can further enhance the results in~\autoref{sec:sample_study}.
In~\autoref{fig:sample_size_acc}, we observe that using more sample data can enhance the deactivating effect on FP, with the ICL performance dropping to 17\% when using 200 and 400 sample data.
Also, using more sample data can slightly improve the ICL performance on TREC and AGnews.
In~\autoref{table:samplesize_ppl} (\autoref{app:results}), we also observe that the PPL change on SST-2 is lower with more sampling.
For instance, the PPL change is 0.339 and 0.399 when the sampling size is 200 and 400, respectively.
Generally, utilizing more sampled data can enhance the ICL controllability of the guarded LLM, but it also demands additional computational resources.

\begin{table}[!t]
\centering
\resizebox{\columnwidth}{!}{
\begin{tabular}{lcccc}
\toprule
 & Target & \multicolumn{3}{c}{Auxiliary} \\
\midrule
\# Sample      & FP & SST-2   & TREC     & AGnews  \\
\midrule
ICLGuard       & 0.31 (-0.49) & 0.80 (-0.13) & 0.68 (-0.02) & 0.82 (-0.08) \\
SFT            & 0.21 (-0.59) & 0.82 (-0.11) & 0.59 (-0.11) & 0.26 (-0.64) \\
DPO            & 0.03 (-0.77) & 0.41 (-0.52) & 0.43 (-0.27) & 0.22 (-0.68) \\
\bottomrule
\end{tabular}
}
\caption{The ICL performance changes with different fine-tuning methods.}
\label{table:baseline_acc}
\end{table}

\mypara{Fine-tuning Methods}
To control the ICL behavior, one straightforward approach is to fine-tune the model to produce an incorrect prediction given the ICL prompt.
Another approach is to fine-tune LLM with reinforcement learning from human feedback, which has become a popular method for aligning models with human values~\cite{CLBMLA17}.
Therefore, we compare \system with Supervised Fine-Tuning (SFT)~\cite{OWJAWMZASRSHKMSAWCLL22} and Direct Preference Optimization (DPO)~\cite{RSMMEF23} approaches on deactivating the ICL for the FP dataset (training details in~\autoref{app:details}).
As shown in~\autoref{table:baseline_acc} and \autoref{table:baseline_ppl}, we report the ICL performance and the model utility.

First, we observe a sharp decline in FP accuracy after applying SFT, showing the success of deactivating the target data.
However, there is a varying degree of decrease in accuracy across other non-target datasets, along with notable changes in the model utility.
We suspect this is due to significant domain shifts or distortions in the model, which could potentially lead to ``concept forgetting.''
In fact, we can consider that training with disable loss is a softer version of SFT, where the model is trained to maximize the log-likelihood of the softly distorted label. 
In contrast, SFT aims to maximize the log-likelihood of the incorrect one-hot label. 
Therefore, training with SFT will likely distort the model significantly, similar to the case where only disable loss is applied, as discussed in~\autoref{sec:loss_study}.
Besides, the result shows that optimizing the model with the one-hot incorrect label will heavily alter it, which demonstrates the need of the soft distorted label.

Second, we find that DPO can deactivate the ICL ability for FP without affecting the model utility. 
However, the ICL performance for other data is affected. 
For instance, the accuracy drops to 41\% for SST-2, 43\% for TREC, and 22\% for AGnews.
This result follows a similar trend observed in~\autoref{sec:loss_study} where only disable and utility losses are applied to train the model. 
It suggests the need to use the maintenance loss to preserve the ICL ability on non-target data.

\begin{table}[!t]
\centering
\resizebox{\columnwidth}{!}{
\begin{tabular}{lcccc}
\toprule
 & Target & \multicolumn{3}{c}{Auxiliary} \\
\midrule
PEFT           & FP (\%) & SST-2 (\%)   & TREC (\%)     & AGnews (\%) \\
\midrule
LoRA     & -0.49 & -0.13 & -0.02 & -0.08 \\ 
PT (8)   & -0.28 & +0.02 & -0.01 & -0.21 \\
PT (16)  & -0.28 & -0.06 & +0.03 & -0.22 \\
\bottomrule
\end{tabular}
}
\caption{The ICL performance with different PEFT methods.
PT = prompt tuning (length).}
\label{table:peft_acc}
\end{table}

\mypara{PEFT}
We study the implications of different PEFT techniques, primarily focusing on LoRA and prompt tuning. 
Our findings, as shown in~\autoref{table:peft_acc}, indicate that prompt tuning can partially deactivate the ICL ability on FP, with a 28\% drop in accuracy.
However, prompt tuning results in low ICL performance on AGnews, achieving only 69\% and 68\% accuracy with 8 and 16-length prompts, respectively.
These results are likely related to the mechanism of prompt tuning, as it is required to append to the input, which could potentially impact the model's performance.
Also, using prompt tuning would produce higher PPL change on SST-2 and TREC, as shown in~\autoref{table:peft_ppl} (\autoref{app:results}).

\mypara{Takeaways}
In this section, we conduct various experiments to explore \system from different perspectives.
We find that \system remains functional with other models or different model sizes.
Additionally, \system is capable of deactivating multiple target data simultaneously, and increasing the number of unique demonstrations and sample data improves the guarded LLM's performance. 
Moreover, we compare the performance of \system to existing fine-tuning methods and show the importance of our loss design.
Last, the use of LoRA generally also yields better results. 
We hope that the findings in this section will assist model owners in selecting the optimal setup when implementing \system on their models.

\subsection{Extension to Generative Task}
\label{sec:gen_study}

\begin{table}[!t]
\centering
\resizebox{\columnwidth}{!}{
\begin{tabular}{lccccc}
\toprule
 & Target & \multicolumn{4}{c}{Auxiliary} \\
\midrule
      & WMT19 (BLEU)  & FP (\%) & SST-2 (\%)   & TREC (\%)     & AGnews (\%) \\
\midrule
Original & 18.24 & 0.80 & 0.93 & 0.70 & 0.90 \\ 
Guarded  & 1.262 & 0.82 & 0.94 & 0.70 & 0.89 \\
\bottomrule
\end{tabular}
}
\caption{The ICL performance of the guarded LLM on deactivating English-to-German translation.}
\label{table:gen_icl}
\end{table}

In the previous section, we have shown that our \system is effective in controlling the ICL behavior for classification tasks.
In practice, the model owner might want to prevent the model from being used for specific generative tasks via ICL, such as generating propaganda.
Although different alignment methods have been proposed to fine-tune the model with human value, we have shown that such methods do not work well for ICL data.
In this section, we investigate the possibility of extending \system for generative tasks.

\mypara{Setup}
To conduct the study for the generative task, we use the WMT19 dataset as an example, which is a collection of parallel corpora used for the 2019 Conference on Machine Translation (WMT19).~\cite{wmt19translate}.
In this study, we assume the model owner wants to prevent users from using the model for English-to-German translation via ICL.
Specifically, we evaluate \system on the LLaMA-13B model using the same ICL setup as~\autoref{sec:exp_setup} with the template ``{sentence} Output:''.
To build the shadow ICL dataset in~\autoref{sec:methodology}, we set the output logit of the first tokens of each ground truth German translation to 0. 
This ensures that the guarded model cannot generate the correct translation or even complete the task.
To evaluate the translation performance, we compute the BLEU score using the NLTK package~\footnote{\url{https://www.nltk.org/}}.

\mypara{Results}
First, we compute the BLEU score for performing WMT19 on LLaMA-13B without ICL. 
Specifically, we use the template ``Translate the sentence to German: {sentence} Output:'' in a zero-shot manner and obtain a BLEU score of 7.17. 
Then, we perform the translation with 16 demonstrations and achieve a BLEU score of 18.24, as shown in \autoref{table:gen_icl}. 
This substantial increase indicates the efficiency and potential of ICL to enhance the model's performance in handling unseen generative tasks.
Next, we evaluate the performance of \system in deactivating ICL for the WMT19 data. 
We observe that the BLEU score drops significantly to 1.262. 
This decrease illustrates the effectiveness of \system in preventing the model from performing the English-to-German translation task.
Meanwhile, the model's utility and ICL performance for other tasks remains comparable to the original performance, as shown in \autoref{table:gen_util}.
The results indicate that \system successfully restricts specific unwanted generative tasks while maintaining the model's capability for other tasks.

\section{Related Works}
\label{sec:related}

\subsection{Large Language Model Safety}

Despite the success of LLMs, they also attract the attention of adversaries.
One major safety concern is adversarial attacks, which are expected to be directed at LLMs.
Zou et al.~\cite{ZWKF23} demonstrated that specific prompts can induce LLMs to produce undesirable behaviors, and these prompts are transferable between different LLMs.
In addition, adversaries might deploy backdoored LLMs designed to produce specific outputs when presented with certain triggers.
BadPrompt~\cite{CXXZY22} generated invisible triggers for each sample in backdoored continuous prompt models.
Besides, a recent concern in the safe and ethical use of LLMs is the phenomenon of ``jailbreaking''~\cite{LDXLZZZZL23}. 
In this, the adversary design input prompts to bypass the LLM's safety mechanisms and generate content that should be restricted.
The potential attacks via these meticulously crafted prompts can be further expanded to ICL.
Therefore, we aim to implement ICL applicability authorization to oversee LLMs.

\subsection{In-Context Learning}

ICL has been discovered in~\cite{BMRSKDNSSAAHKHCRZWWHCSLGCCBMRSA20} that can perform tasks without any parameter updates.
Since then, various studies have been proposed to understand the mechanism behind it.
Oswald et al.~\cite{ONRSMZV23} suggested that ICL in transformers approximates gradient-based few-shot learning during its forward pass. 
Conversely, Xie et al.~\cite{XRLM22} interpreted ICL as implicit Bayesian inference, proposing that ICL emerges when the pre-training distribution blends with the hidden Markov models. 
Besides, Min et al.~\cite{MLHALHZ22} explored various factors, such as the use of ground-truth labels, that can affect ICL performance.
In addition, Dai et al.~\cite{DSDHMSW23} demonstrated similarities between in-context learning and explicit fine-tuning through measuring with different metrics on different datasets.
Prior studies mainly focus on understanding ICL and how to improve it.
Instead, we explore the possibility of controlling ICL behavior.

\subsection{In-Context Learning Safety}

As the popularity of ICL surges, researchers have increasingly focused on its safety implications.
Kandpal et al.~\cite{KJTC23} introduced backdoors to the LLM using ICL prompts, and the backdoored model exhibited malicious behavior when presented with triggered inputs.
Duan et al.~\cite{DDYPB23} studied privacy leakage on LLMs using membership inference attacks.
They assume that some LLMs contain pre-defined ICL prompts for the downstream task, which might contain sensitive data.
They argue that deploying prompted models presents a significant privacy risk.
Panda et al.~\cite{PWWM23} proposed Differentially Private In-context Learning (DP-ICL) for privatizing ICL tasks.
DP-ICL produces differentially private responses with a set of LLM responses and a noisy consensus mechanism.
In this paper, we pivot to the safety concerns of ICL from a unique angle, emphasizing the control of model behavior in response to ICL prompts with different data.

\subsection{Model Authorization}

Model authorization was initially proposed to protect model intellectual property.
Alam et al.~\cite{ASMK20} encrypted network parameters using a secret key.
Chakraborty et al.~\cite{CMS20} generated a secret key from the hardware fingerprints of a specific device, and only users with that device are allowed to load and utilize the model.
On the other hand, Wang et al.~\cite{WXXWZ22} introduced applicability authorization for data-centric protection.
Instead of specifying who can access the model, this approach dictates which data the model can process.
For instance, a model designed for the MNIST dataset might underperform when presented with the USPS dataset.
We build upon the concept of applicability authorization, extending its principles to ICL. 
Our objective is to regulate ICL behavior by determining what data the model can process using ICL.

\section{Conclusion}
\label{sec:discussion}

This paper introduces the first fine-tuning framework, \system, designed to regulate ICL behavior in LLMs. 
Drawing inspiration from the "applicability authorization" concept tailored for LLMs, \system employs three distinct loss functions to optimize the PEFT module to safeguard the LLM. 
Our approach aims to deactivate the ICL capability when presented with target data while preserving its function for non-target data. 
Additionally, the regular functionalities of the LLM remain intact for all data. 
Experimental findings highlight that \system effectively protects the LLM from potential misuse. 
For example, when applied to the LLaMA-13B model, \system reduces its ICL ability on FP to 31\%, yet it sustains strong ICL performance on benchmark datasets such as SST-2, TREC, and AGnews. 
Further, our research investigates the individual impact of each loss function on modulating ICL behaviors and explores advanced techniques for creating data sets that extend beyond the target domain.

In the later stages of our study, we explore the resilience of \system against adaptive attacks. 
This involves simulating scenarios where malicious users might adjust their prompting to bypass the applicability authorization mechanisms implemented by \system. 
By assessing these potential vulnerabilities, we can better understand the robustness of our system.
We also delve into the effect of various setups that can influence the framework's effectiveness in regulating ICL behavior within LLMs.
We find that \system is still effective on different models and different model sizes and is capable of deactivating multiple target data simultaneously.
Also, \system has better performance in controlling the ICL performance compared to existing fine-tuning methods.
Last, we are able to adapt \system to control the generative task under ICL.
We hope that these findings can offer valuable insights for model owners in selecting the optimal setup when implementing \system on their models.

\begin{small}
\bibliographystyle{plain}
\bibliography{normal_generated_py3}
\end{small}

\newpage
\appendix

\section{Experiment Details}
\label{app:appendix}

\subsection{Prompt Templates}
\label{app:template}

For each task, we use the minimum template and default label set.
For FP, we use three extra templates for the study of the adaptive attack.
All prompts and labels are presented in~\autoref{table:temp_set} and~\autoref{table:label_set}.

\subsection{Baseline Performance}
\label{app:baseline}

For the model and size study, we provide the original ICL performance and utility for OPT-13B, Cerebras-13B, LLaMA-7B, and LLaMA-30B in~\autoref{table:opt_baseline}, \autoref{table:cerebras_baseline}, \autoref{table:7b_baseline}, and \autoref{table:30b_baseline}, respectively.

\subsection{More Results}
\label{app:results}

We report the utility performance for~\autoref{sec:further_study}, including different models, different LLaMA model sizes, different numbers of unique demonstrations, different sampling sizes, and different PEFT methods in~\autoref{table:model_ppl}, \autoref{table:modelsize_ppl}, \autoref{table:uniquedemo_ppl}, \autoref{table:samplesize_ppl}, and \autoref{table:peft_ppl}.

\subsection{Training Details}
\label{app:details}

For SFT and DPO, we use the same setup as~\autoref{sec:exp_setup} for fine-tuning the LoRA.
We set the rank to 8, alpha to 32, and the dropout rate to 0.1. 
In addition, we use a batch size of 4 and a learning rate of 1 x $10^{-4}$ with the Adam optimizer.
For SFT, we randomly select an incorrect label as the continuous on the target ICL prompt, and truncate the input length to 400.
For DPO, we set the ground label as the rejected answer, and randomly select an incorrect label as the chosen answer.
In addition, we quantize the model to 4 bits.
We implement the training using Huggingface's TRL package~\footnote{\url{https://github.com/huggingface/trl}}.

\begin{table}[!t]
\centering
\resizebox{\columnwidth}{!}{
\begin{tabular}{ll}
\toprule
 \bf Template \# & \bf Example \\
\midrule
0 & \multicolumn{1}{l}{\begin{tabular}[c]{@{}l@{}}\textless{}Sentence\textgreater \\ Label: \textless{}Label\textgreater{}\end{tabular}} \\
\midrule
1 & \multicolumn{1}{l}{\begin{tabular}[c]{@{}l@{}}\textless{}Sentence\textgreater \\ Sentiment: \textless{}Label\textgreater{}\end{tabular}} \\
\midrule
2 & \multicolumn{1}{l}{\begin{tabular}[c]{@{}l@{}}\textless{}Sentence\textgreater \\ The sentiment is \textless{}Label\textgreater{}\end{tabular}} \\
\midrule
3 & \multicolumn{1}{l}{\begin{tabular}[c]{@{}l@{}}\textless{}Sentence\textgreater \\ The sentiment of the text is \textless{}Label\textgreater{}\end{tabular}} \\
\bottomrule
\end{tabular}
}
\caption{List of templates taken from~\cite{PGCC23}.}
\label{table:temp_set}
\end{table}

\begin{table}[!t]
\centering
\resizebox{\columnwidth}{!}{
\begin{tabular}{lll}
\toprule
\bf Dataset & \bf Label \# & \\
\midrule
\multirow{2}{*}{FP} & 1 & negative/neutral/positive \\
 & 2 & terrible/neutral/good \\
\midrule
SST-2 & 1 & terrible/good/neutral \\
\midrule
TREC & 1 & \multicolumn{1}{l}{\begin{tabular}[c]{@{}l@{}}expression/entity/description/\\ human/location/number\end{tabular}} \\
\midrule
AGnews & 1 & World/Sports/Business/Science \\
\bottomrule
\end{tabular}
}
\caption{List of default labels for each dataset.}
\label{table:label_set}
\end{table}

\begin{table}[!t]
\centering
\begin{tabular}{lcc}
\toprule
Dataset & Accuracy (\%) & PPL     \\
\midrule
FP      & 0.80  & 31.42\\
SST-2   & 0.93  & 405.0\\
TREC    & 0.70  & 51.92\\
AGnews  & 0.90  & 11.25\\
Lambada  & -     & 27.64\\
\bottomrule
\end{tabular}
\caption{The original ICL performance and utility of LLaMA-13B for each dataset.}
\label{table:baseline}
\end{table}

\begin{table}[!t]
\centering
\begin{tabular}{lcc}
\toprule
Dataset & Accuracy (\%) & PPL     \\
\midrule
FP      & 0.73  & 98.01   \\
SST-2   & 0.93  & 443.3   \\
TREC    & 0.56  & 129.1   \\
AGnews  & 0.78  & 25.81   \\
Lambada  & -     & 37.41   \\
\bottomrule
\end{tabular}
\caption{The original ICL performance and utility of OPT-13B for each dataset.}
\label{table:opt_baseline}
\end{table}

\begin{table}[!t]
\centering
\begin{tabular}{lcc}
\toprule
Dataset & Accuracy (\%) & PPL     \\
\midrule
FP      & 0.55  & 100.3   \\
SST-2   & 0.89  & 350.1  \\
TREC    & 0.41  & 345.1   \\
AGnews  & 0.77  & 33.81   \\
Lambada  & -     & 44.46   \\
\bottomrule
\end{tabular}
\caption{The original ICL performance and utility of Cerebras-13B for each dataset.}
\label{table:cerebras_baseline}
\end{table}

\begin{table}[!t]
\centering
\begin{tabular}{lcc}
\toprule
Dataset & Accuracy (\%) & PPL     \\
\midrule
FP      & 0.77  & 32.28   \\
SST-2   & 0.95  & 572.9  \\
TREC    & 0.80  & 52.61   \\
AGnews  & 0.89  & 12.05   \\
Lambada  & -     & 28.39   \\
\bottomrule
\end{tabular}
\caption{The original ICL performance and utility of LLaMA-7B for each dataset.}
\label{table:7b_baseline}
\end{table}

\begin{table}[!t]
\centering
\begin{tabular}{lcc}
\toprule
Dataset & Accuracy (\%) & PPL     \\
\midrule
FP      & 0.83  & 28.75   \\
SST-2   & 0.95  & 509.9  \\
TREC    & 0.92  & 51.18   \\
AGnews  & 0.91  & 10.66   \\
Lambada  & -     & 26.45   \\
\bottomrule
\end{tabular}
\caption{The original ICL performance and utility of LLaMA-30B for each dataset.}
\label{table:30b_baseline}
\end{table}

\begin{table*}[!t]
\centering
\begin{tabular}{lccccc}
\toprule
Model       & FP & SST-2    & TREC  & AGnews & Lambada \\
\midrule
LLaMA-13B    & 31.44 (+0.020) & 402.8 (-2.164) & 51.95 (+0.024) & 11.24 (-0.004) & 27.63 (-0.012) \\
OPT-13B      & 87.12 (-10.89) & 417.9 (-25.35) & 132.3 (+3.124) & 25.62 (-0.186) & 39.14 (+1.728) \\
Cerebras-13B & 100.2 (-0.077) & 349.8 (-0.293) & 344.7 (-0.417) & 33.80 (-0.014) & 44.46 (-0.003) \\
\bottomrule
\end{tabular}
\caption{The utility and changes with different models.}
\label{table:model_ppl}
\end{table*}

\begin{table*}[!t]
\centering
\begin{tabular}{lccccc}
\toprule
Model       & FP  & SST-2  & TREC  & AGnews & Lambada \\
\midrule
7B     & 32.27 (-0.011) & 571.0 (-1.924) & 52.86 (+0.252) & 12.05 (+0.006) & 28.40 (+0.004) \\
13B    & 31.44 (+0.020) & 402.8 (-2.164) & 51.95 (+0.024) & 11.24 (-0.004) & 27.63 (-0.012) \\ 
30B    & 28.74 (-0.006) & 511.4 (+1.538) & 51.19 (+0.008) & 10.66 (+0.007) & 26.46 (+0.014) \\
\bottomrule
\end{tabular}
\caption{The utility and changes with different sizes of LLaMA model.}
\label{table:modelsize_ppl}
\end{table*}

\begin{table*}[!t]
\centering
\begin{tabular}{lccccc}
\toprule
Target          & FP     & SST-2  & TREC   & AGnews & Lambada \\
\midrule
FP              & +0.020 & -2.164 & +0.024 & -0.004 & -0.012 \\ 
FP+SST-2        & -0.006 & -2.189 & +0.101 & -0.005 & +0.049 \\
FP+SST-2+AGnews & +0.014 & +0.437 & +0.212 & +0.006 & +0.041 \\
\bottomrule
\end{tabular}
\caption{The utility changes with different target settings on the LLaMA model.}
\label{table:multi_ppl}
\end{table*}

\begin{table*}[!t]
\centering
\resizebox{\columnwidth}{!}{
\begin{tabular}{lccccc}
\toprule
 & Target & \multicolumn{3}{c}{Auxiliary} \\
\midrule
\# Unique              & FP & SST-2   & TREC     & AGnews & Lambada \\
\midrule
10   & +0.016 & -2.168 & +0.104 & +0.005 & +0.013 \\
20   & +0.010 & -1.955 & +0.074 & -0.004 & +0.008 \\
40   & +0.020 & -2.164 & +0.024 & -0.004 & -0.012 \\ 
80   & +0.022 & +0.509 & +0.068 & -0.006 & +0.034 \\ 
\bottomrule
\end{tabular}
}
\caption{The utility changes with different numbers of unique demonstrations.}
\label{table:uniquedemo_ppl}
\end{table*}

\begin{table*}[!t]
\centering
\resizebox{\columnwidth}{!}{
\begin{tabular}{lccccc}
\toprule
 & Target & \multicolumn{3}{c}{Auxiliary} \\
\midrule
\# Sample              & FP & SST-2   & TREC     & AGnews & Lambada \\
\midrule
100   & +0.020 & -2.164 & +0.024 & -0.004 & -0.012 \\ 
200   & +0.004 & +0.339 & +0.006 & +0.063 & -0.003 \\
400   & -0.017 & +0.399 & -0.051 & +0.001 & +0.002\\
\bottomrule
\end{tabular}
}
\caption{The utility changes with different sampling sizes.}
\label{table:samplesize_ppl}
\end{table*}

\begin{table*}[!t]
\centering
\resizebox{\columnwidth}{!}{
\begin{tabular}{lccccc}
\toprule
 & Target & \multicolumn{3}{c}{Auxiliary} \\
\midrule
\# Sample      & FP & SST-2   & TREC     & AGnews & Lambada \\
\midrule
ICLGuard       & +0.020 & -2.164 & +0.024 & -0.004 & -0.012 \\ 
SFT            & +608.5 & +881.1 & inf    & +313.2 & inf   \\
DPO            & +0.415 & +19.20 & +0.504 & +0.086 & -0.185\\
\bottomrule
\end{tabular}
}
\caption{The utility changes with different fine-tuning methods.}
\label{table:baseline_ppl}
\end{table*}

\begin{table*}[!t]
\centering
\resizebox{\columnwidth}{!}{
\begin{tabular}{lccccc}
\toprule
 & Target & \multicolumn{3}{c}{Auxiliary} \\
\midrule
PEFT    & FP     & SST-2  & TREC   & AGnews & Lambada \\
\midrule
LoRA    & +0.020 & -2.164 & +0.024 & -0.004 & -0.012 \\ 
PT (8)  & +0.379 & -292.1 & +5.888 & +0.063 & +0.698 \\
PT (16) & -0.167 & -256.0 & -4.399 & +0.001 & +0.059 \\
\bottomrule
\end{tabular}
}
\caption{The utility changes with different PEFT methods.
PT = prompt tuning.}
\label{table:peft_ppl}
\end{table*}

\begin{table*}[!t]
\centering
\resizebox{\columnwidth}{!}{
\begin{tabular}{lccccc}
\midrule
    & FP     & SST-2  & TREC   & AGnews & Lambada \\
\midrule
Original & 31.42 & 405.0 & 51.92 & 11.25 & 27.64 \\
Guarded  & 31.44 & 405.4 & 52.14 & 11.25 & 27.64 \\
\bottomrule
\end{tabular}
}
\caption{The utility of the guarded LLM on deactivating English-to-German translation.}
\label{table:gen_util}
\end{table*}

\end{document}